\documentclass[11pt]{elsarticle}
\usepackage{lineno,hyperref}
\modulolinenumbers[5]
\usepackage{natbib}
\usepackage{stmaryrd}
\usepackage{subfigure}
\usepackage{color} 
\setcitestyle{authoryear}
\usepackage{amsmath}
\usepackage{bm}
\usepackage{bbm}
\usepackage{mathrsfs}
\usepackage{graphicx}
\usepackage{caption}
\usepackage{epsfig}
\usepackage{amsfonts}
\usepackage{amsmath}
\usepackage{graphicx}
\usepackage{psfrag}
\usepackage{placeins}
\graphicspath{{./figures/}}
\def\eps{\varepsilon}
\def\evat{\bigg |}

\def\Eshelby{\boldsymbol{\mathcal{E}}}

\def\half{{\textstyle{1 \over 2}}}

\def\Bnabla{\nabla}

\def\ljmp{{\lbrack\!\lbrack}}
\def\rjmp{{\rbrack\!\rbrack}}

\def\BTheta{\mbox{\boldmath$\Theta$}}

\def\BXi{\mbox{\boldmath$\Xi$}}

\def\Bkappa{\mbox{\boldmath$\kappa$}}

\def\Bphi{\mbox{\boldmath$\phi$}}
\def\Bchi{\mbox{\boldmath$\chi$}}

\def\iso{\mathbbm{1}}

\def\bB{\mbox{\boldmath$ B$}}
\def\bC{\mbox{\boldmath$ C$}}
\def\bD{\mbox{\boldmath$ D$}}
\def\bE{\mbox{\boldmath$ E$}}
\def\bF{\mbox{\boldmath$ F$}}

\def\bM{\mbox{\boldmath$ M$}}
\def\bN{\mbox{\boldmath$ N$}}

\def\bP{\mbox{\boldmath$ P$}}

\def\bT{\mbox{\boldmath$ T$}}
\def\bU{\mbox{\boldmath$ U$}}
\def\bV{\mbox{\boldmath$ V$}}
\def\bW{\mbox{\boldmath$ W$}}
\def\bX{\mbox{\boldmath$ X$}}

\def\ba{\mbox{\boldmath$ a$}}
\def\bb{\mbox{\boldmath$ b$}}
\def\bc{\mbox{\boldmath$ c$}}

\def\be{\mbox{\boldmath$ e$}}
\def\bf{\mbox{\boldmath$ f$}}

\def\br{\mbox{\boldmath$ r$}}

\def\bu{\mbox{\boldmath$ u$}}
\def\bv{\mbox{\boldmath$ v$}}
\def\bw{\mbox{\boldmath$ w$}}
\def\bx{\mbox{\boldmath$ x$}}

\begin{document}
\title{A variational treatment of material configurations with application to interface motion and microstructural evolution}
\author{Gregory H. Teichert}\address{Mechanical Engineering, University of Michigan}
\author{Shiva Rudraraju} \address{Mechanical Engineering, University of Michigan}
\author{Krishna Garikipati\corref{mycorrespondingauthor}}\address{Mechanical Engineering, \& Mathematics, University of Michigan}
\cortext[mycorrespondingauthor]{Corresponding Author}
\ead{krishna@umich.edu}
\begin{abstract} We present a unified variational treatment of evolving configurations in crystalline solids with microstructure. The crux of our treatment lies in the introduction of a vector configurational field. This field lies in the material, or configurational, manifold, in contrast with the traditional displacement field, which we regard as lying in the spatial manifold. We identify two distinct cases which describe (a) problems in which the configurational field's evolution is localized to a mathematically sharp interface, and (b) those in which the configurational field's evolution can extend throughout the volume. The first case is suitable for describing incoherent phase interfaces in polycrystalline solids, and the latter is useful for describing smooth changes in crystal structure and naturally incorporates coherent (diffuse) phase interfaces. These descriptions also lead to parameterizations of the free energies for the two cases, from which variational treatments can be developed and equilibrium conditions obtained. For sharp interfaces that are out-of-equilibrium, the second law of thermodynamics furnishes restrictions on the kinetic law for the interface velocity. The class of problems in which the material undergoes configurational changes between distinct, stable crystal structures are characterized by free energy density functions that are non-convex with respect to configurational strain. For physically meaningful solutions and mathematical well-posedness, it becomes necessary to incorporate interfacial energy. This we have done by introducing a configurational strain gradient dependence in the free energy density function following ideas laid out by Toupin (\emph{Arch. Rat. Mech. Anal.}, \textbf{11}, 1962, 385-414). The variational treatment leads to a system of partial differential equations governing the configuration that is coupled with the traditional equations of nonlinear elasticity. The coupled system of equations governs the configurational change in crystal structure, and elastic deformation driven by elastic, Eshelbian, and configurational stresses. Numerical examples are presented to demonstrate interface motion as well as evolving microstructures of crystal structures.
\end{abstract}

\maketitle
\section*{Keywords} Free energy; thermodynamic driving force; gradient regularization; configurational stress; spline basis functions.

\section{Introduction}

We present a variational treatment of evolving configurations in solids. Of interest to us are problems in which a kinematic field can be identified, which describes the essential aspects of the material's configuration, while another distinct field, the displacement, furnishes the kinematics necessary for representing the nonlinear elastic response. Such a separation is possible upon a suitable definition of configurations for the cases at hand. A series of mathematical steps can then follow: The total free energy can then be written as a functional of both the configurational and the displacement fields. With it, we can seek equilibrium states that render the free energy stationary with respect to both fields. The corresponding Euler-Lagrange equations governing the configurational and displacement fields can be solved. The motivation from physics comes of asking whether a solid under load can seek to reach equilibrium by varying some configurational degree of freedom that can be identified as being distinct from the displacement field.

The somewhat abstract arguments laid out above have relevance to crystalline solids that undergo phase transformations coupled with elastic deformation: In a classical continuum setting, the elastic deformation is  obtained from the displacement, which is the only kinematic field. No phenomena are sought to be modeled, other than the mapping of the reference to current placements. In this setting, the reference and material placements coincide, and most importantly, they are fixed. In contrast stands any phenomenon in which, the material configuration evolves from a \emph{reference material configuration}, and can be represented, on a physical basis, by a configurational field that is distinct from the displacement field. Here, we are concerned with two specific examples: (a) In a multi-phase solid where phase change occurs at interphase interfaces, the configurational field would represent interface migration. The phase, and therefore the crystal structure at a material point will change if the interface migrates through that point. This causes a change in the material configuration of the point. Since the crystal structure (material configuration) changes across the mathematically sharp interface, the latter is incoherent. (b) Alternately, in a multi-phase solid, the crystal structure may change smoothly from one phase to another over an interface that has finite width, rather than being mathematically sharp. In this case also, the  material configuration evolves with the crystal structure. Clearly, this would be a case of coherent interphase interfaces. Here too, the configurational field would represent the crystal structure at any point in the solid as a map from some well-defined, reference material configuration.

In each of cases a and b, with the evolved material configuration determined as above, the displacement field can be defined as the point-to-point map from this configuration to the current/deformed placement that lies in the spatial manifold. An elastic deformation can then be identified from this displacement field.

With two distinct kinematic fields thus identified, the response of the solid can be described by parameterizing the free energy functional in terms of these two fields. The imposition of equilibrium as the conditions of stationarity under variations on the configurational and displacement fields reveals two sets of Euler-Lagrange equations. As expected, one set contains the standard partial differential equations of elasticity. The second set is novel, and consists of partial differential equations and accompanying boundary conditions that involve the conventional elastic stress, the Eshelby stress, as well as a distinct configurational stress. 

The treatment of a configurational force, distinct from standard, Newtonian, forces acting on imperfections in a crystal lattice was given by \citet{Eshelby1951}, building off work from the late nineteenth century \citep{Burton1892,Larmor1897}. The last two decades have seen a resurgence in the literature on configurational forces. Some of the theoretical underpinnings can be found in  \cite{Gurtin2000,Maugin1995,KienzlerHerrmann1997,Steinmann2002,Maugin2011} and \citet{VuSteinmann2012}. Applications have also been developed, such as to finite element discretization \citep{MullerMaugin2002}, to the dynamics of defects \citep{AcharyaFressengeas2012}, to spatial and material covariant balance laws \citep{Yavarietal2006} for modeling elastic inclusions \citep{YavariGoriely2013}, and to fracture mechanics \citep{DenzerMenzel2014}, Configurational force equations can be derived in the setting of classical balance laws, or, with appropriate assumptions, within a variational framework. \citet{Gurtin2000} regards configurational forces as fundamental quantities in continuum physics, analogous to standard forces. On that premise, he regards configurational balance laws as the corresponding, fundamental laws that must exist in order to govern these forces. This has led to a debate on whether new physics is posited by the introduction of configurational forces \citep{Maugin2011,PodioGuidugli2002}. The work we present here lies within a variational setting, and circumvents this debate by relying on the (perhaps) more accepted notion of equilibrium to arrive at balance laws as Euler-Lagrange equations of free energy functionals. The resulting partial differential equation for configurational equilibrium also arises in \citet{Gurtin2000}, and in \citet{Maugin2011}, where it has been called the fully material equilibrium equation.

We consider first the problem of configurational changes taking place at a sharp, migrating interface between two solid material phases. We show that the variational method produces a partial differential equation of configurational equilibrium in addition to the standard partial differential equation of elasticity. Assuming satisfaction of quasi-static elastic equilibrium, the partial differential equation for configurational equilibrium is identically satisfied everywhere except on the interface itself. There, it takes the form of a jump condition, which also vanishes if equilibrium is satisfied at the interface. However, it is of interest to consider solids that are far from equilibrium, and therefore have migrating interfaces. Then, the second law of thermodynamics provides guidance for choosing a sufficient form for the interface velocity. We adopt the well-known and widely-used level set method \citep{OsherSethian1988} to track the interface's motion based on this velocity. Here, we list just a few of a vast number of level set applications: \citet{BarthSethian1998} modeled an isotropic etching process with a constant velocity and a directional etching process with a velocity dependent on the interface orientation. \citet{MacklinLowengrub2006} modeled tumor growth with a curvature dependent velocity. The velocity in oxidation problems modeled by \citet{Raoetal2000,RaoHughes2000}, \citet{GarikipatiRao2001} is based on material composition. Finally, we note that \citet{KalpakidesArvanitakis2009} used a velocity based on configurational forces to model ferroelastic materials, although it is arrived at differently than in the present work.

We next turn to the problem of smoothly varying configurational changes in crystal structure that occur over interfaces of finite width. The smoothness implies that the configurational change extends over finite sub-volumes and transforms the crystal structure from the parent to the daughter phases. Therefore, it is in contrast to the case of phase transformation only at a migrating sharp interface. The configurational change of the crystal structure over the volume suggests that there is a contribution to the free energy density function, which is associated with this configurational field. The variational treatment based on stationarity of the free energy functional leads to a partial differential equation for configurational balance that holds throughout the volume of the crystalline solid. There is therefore a fundamental difference in the form of the governing equations from that for phase changes that occur only at a sharp interface. The variational setup, however, is similar in both problems. Such a consideration of configurational change that occurs over the volume of a material was attempted, albeit in a limited manner, by \citet{Garikipatietal2006} in the context of remodeling in biology. Since the parent and daughter crystal structures are equilibrium structures under suitable conditions, the free energy density function must exhibit local minima in configurational tangent space corresponding to these structures. The free energy density function is therefore non-convex and admits microstructures, thus placing the problem in a class that has spawned a rich mathematical literature including \cite{BhattacharyaKohn1997,Muller1999,Bhattacharyaetal2004}. It also is well-known that the non-convex free energy density functions must be enhanced by terms that penalize gradients in the tangent maps of the configurational variables for mathematical well-posedness and physically meaningful solutions \citep{BallCrooks2011}. Such considerations were accounted for by \cite{Rudrarajuetal2014,Rudrarajuetal2015}, who treated non-convex free energy density-driven microstructure formation in nonlinear gradient elasticity following \citet{Toupin1962}. We have extended this treatment to the configurational field in this communication.

Our treatment begins with consideration of the problem where the configurational change is restricted to a sharp interface in Section \ref{sec:sharpinterface}, and then moves on to the problem of smoothly varying configurational change over a diffuse interface in Section \ref{sec:diffuseinterface}. The treatment is illustrated by numerical examples in both sections. Concluding remarks appear in Section \ref{sec:conclusion}.

\section{Configurational change restricted to a sharp interface}
\label{sec:sharpinterface}
      \begin{figure}[htb]
        \centering
        \includegraphics[width=\textwidth]{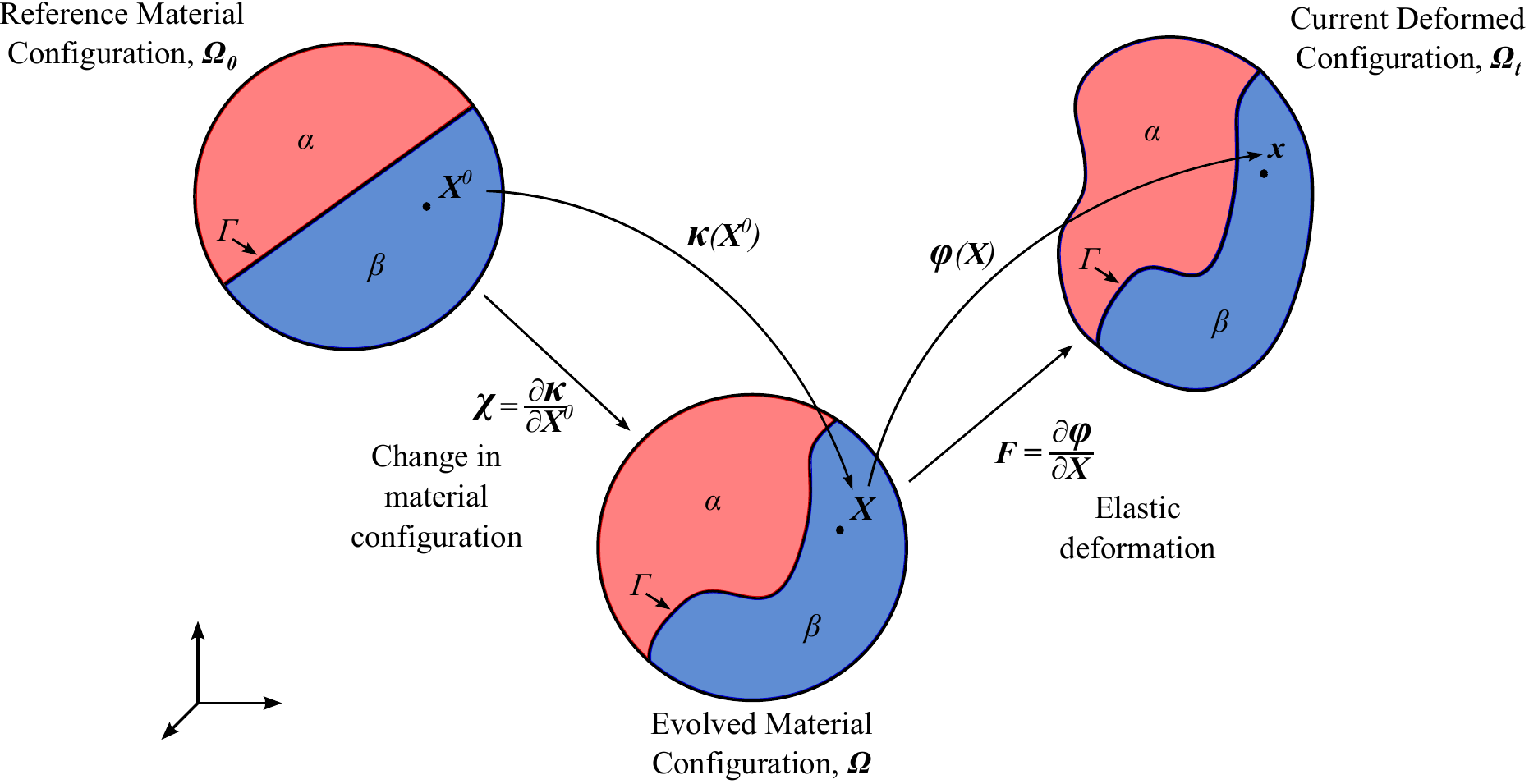}
        \caption{Kinematics of the configurational changes and elastic deformations with a sharp interface.}
	\label{fig:sharp}
      \end{figure}

Consider a body that is an open set $\Omega \subset \mathbb{R}^3$ with two open subsets, $\Omega_\alpha$ and $\Omega_\beta$ consisting of phases $\alpha$ and $\beta$, respectively, that meet at a sharp interface $\Gamma \subset \Omega$. Here $\Gamma$ is a 2-manifold that can be constructed as a mapping $\Gamma: \mathbb{R}^2 \mapsto \mathbb{R}^3$ (see Figure~\ref{fig:sharp}). Let
\begin{align}
\overline{\Omega}  &= \overline{\Omega_\alpha \cup \Omega_\beta} \nonumber \\
\overline{\Gamma} &= \partial \Omega_\alpha \cap \partial \Omega_\beta \nonumber \\
\partial \Omega &= \partial \Omega_\alpha \cup \partial \Omega_\beta \setminus \Gamma
\end{align}
We denote the traditional traction boundary, where external, standard tractions can be imposed, by $\partial \Omega^\mathrm{S}_T$, while the traction-like boundary related to changes in material configuration is $\partial \Omega^\mathrm{M}_T$. This description is of the body in some material configuration.

We suppose that the body has arrived at the above configuration by undergoing a phase transformation from the reference material configuration $\Omega_0$, characterized by motion of the interface from its reference configuration $\Gamma_0$ to $\Gamma$ in configurational space. The reference material configurations of $\Omega_\alpha$ and $\Omega_\beta$ are, respectively, $\Omega_{\alpha_0}$ and $\Omega_{\beta_0}$. The point-to-point map $\Bkappa$ is from the reference material configuration $\Omega_0$ onto the evolved material configuration $\Omega$.
\begin{align}
\bX &= \Bkappa(\bX^0) =  \bX^0 + \bU \label{eqn:X}\\
\Bchi &= \frac{\partial \Bkappa}{\partial \bX^0} = \iso +  \frac{\partial \bU}{\partial \bX^0}
\label{eqn:chi}
\end{align}
Here, $\bX^0$ and $\bX$ are the reference and evolved values of the configurational field. We will refer to $\bU$ as the configurational displacement, since it describes the change in material configuration through the evolution of the phases and their interface.

The elastic deformation of the body is described in the standard manner based on the mapping $\Bphi:\Omega \mapsto \Omega_t$, defined as:
\begin{equation}
\bx = \Bphi(\bX) =  \bX + \bu
\end{equation}
where $\bu$ is the standard displacement field. The elastic deformation gradient is then defined as
\begin{equation}
\bF = \frac{\partial\Bphi}{\partial\bX^0}\frac{\partial\bX^0}{\partial\bX} = \frac{\partial \Bphi}{\partial \bX} = \iso +  \frac{\partial \bu}{\partial \bX}
\end{equation}
These variables can also be written in terms of $\bX^0$, as follows:
\begin{align}
\bx &= \Bphi(\Bkappa(\bX^0)) =  \bX^0 + \bU + \bu\\
\bF
 &= \left(\iso + \frac{\partial \bU}{\partial \bX^0} + \frac{\partial \bu}{\partial \bX^0}\right)
\left( \iso + \frac{\partial \bU}{\partial \bX^0} \right)^{-1}
\label{eqn:F}
\end{align}
We emphasize the distinctions between the configurational map, $\Bkappa$ and the deformation map $\Bphi$, their respective tangent maps $\Bchi$ and $\bF$, and the associated configurational and standard displacement fields $\bU$ and $\bu$. 

The Gibbs free energy of the system is given by the following functional defined over $\Omega$, where $\psi^\alpha(\bF,\bX)$ and $\psi^\beta(\bF,\bX)$ are the strain energy density functions for the two phases:
\begin{align}
\begin{aligned}[b]
\Pi[\bu;\bU] 
&=
\int \limits_{{\Omega_\alpha}}\psi^\alpha(\bF,\bX)\,\mathrm{d}V
+ \int \limits_{{\Omega_\beta}}\psi^\beta(\bF,\bX)\,\mathrm{d}V\\
&\phantom{=}
 - \int \limits_{{\Omega}} \bf(\bX) \cdot \bu \, \mathrm{d}V
 -  \int \limits_{\partial \Omega^\mathrm{S}_T}  \bT \cdot \bu\, \mathrm{d}S
\end{aligned}
\end{align}
We perform a change of variables to define the functional over $\Omega_0$, noting that constancy of the traction loading during configurational change implies that $\bT \mathrm{d}S = \bT^0\mathrm{d}S_0$:
\begin{align}
&
\begin{aligned}[b]
\Pi[\bu;\bU]
&=\int \limits_{{\Omega}_{\alpha_0}}\psi^\alpha(\bF,\Bkappa(\bX^0))\det\Bchi\,\mathrm{d}V_0
+ \int \limits_{{\Omega}_{\beta_0}}\psi^\beta(\bF,\Bkappa(\bX^0))\det\Bchi\,\mathrm{d}V_0\\
&\phantom{=}
- \int \limits_{{\Omega}_0}\bf(\Bkappa(\bX^0)) \cdot \bu \det\Bchi\, \mathrm{d}V_0
 -  \int \limits_{{\partial \Omega^\mathrm{S}_{T_0}}} \bT^0 \cdot \bu \, \mathrm{d}S_0
\end{aligned}
\end{align}
\subsection{Variational formulation}
 We consider variations on the configurational displacement, $\bU^\eps :=\bU + \eps \bW$, and on the standard displacement, $\bu^\eps := \bu + \eps \bw$. We compute the first variation using the functional defined over the fixed, reference material configuration $\Omega_0$. A complete derivation is presented in \ref{sec:appa}. Note that we express the two integrals $\int_{\Omega_\alpha}\psi^\alpha$ and $\int_{\Omega_\beta}\psi^\beta$ with a single integral $\int_{\Omega}\psi$.

\begin{align}
\begin{aligned}[b]
\frac{\mathrm{d}}{\mathrm{d}\eps} \Pi[\bu^\eps;\bU^\eps] \evat_{\eps=0} &=
\frac{\mathrm{d}}{\mathrm{d}\eps} \bigg\{ 
\int \limits_{{\Omega}_0}\psi(\bF^\eps,\Bkappa^\eps(\bX^0))\det\Bchi^\eps\,\mathrm{d}V_0\\
&\phantom{=}
 - \int \limits_{{\Omega}_0} \bf(\Bkappa^\eps(\bX^0)) \cdot \bu^\eps\det\Bchi^\eps \, \mathrm{d}V_0\\
&\phantom{=}
 -  \int \limits_{{\partial \Omega^\mathrm{S}_{T_0}}} \bT^0 \cdot \bu^\eps \, \mathrm{d}S_0 \bigg\}
\evat_{\eps=0}
\end{aligned}
\label{eqn:PI_eps}
\end{align}
At equilibrium, the first variation of the Gibbs free energy is zero: $[\mathrm{d}\Pi/\mathrm{d}\eps]_{\eps = 0} = 0$.
The first variations of $\bF$ and $\det \Bchi$ are derived from equations (\ref{eqn:F}) and (\ref{eqn:chi}), respectively:
\begin{align}
\frac{d \bF^\eps}{d \eps}
&= \left[ \frac{\partial \bw}{\partial \bX^0}
 + \left(\mathbbm{1} - \bF^\eps \right)
\frac{\partial \bW}{\partial \bX^0}
\right] {\Bchi^\eps}^{-1}
\label{eqn:F_eps}\\
\frac{d \det\Bchi^\eps}{d \eps} 
&=  \iso:\left(\frac{\partial \bW}{\partial \bX^0} {\Bchi^\eps}^{-1}\right)\det\Bchi^\eps
\label{eqn:det_chi_eps}
\end{align}

Substituting (\ref{eqn:F_eps}) and (\ref{eqn:det_chi_eps}) into (\ref{eqn:PI_eps}) defining the first Piola-Kirchhoff stress as $\partial \psi/\partial\bF = \bP$, and using $d\Bkappa^\eps/d\eps = \bW$ and $d\bu^\eps/d\eps = \bw$ gives
\begin{align}
\begin{aligned}[b]
0 &=
\int \limits_{{\Omega_0}}
\bP: \left(\frac{\partial \bw}{\partial \bX^0}\Bchi^{-1}\right)
\det\Bchi\,\mathrm{d}V_0
 - \int \limits_{{\Omega_0}} (\bf\cdot\bw)\det\Bchi \, \mathrm{d}V_0\\
&\phantom{=}
+ \int \limits_{{\Omega_0}}
\left(\bP - \left(\bf\cdot\bu\right)\iso + \Eshelby \right):
\left(\frac{\partial \bW}{\partial \bX^0}\Bchi^{-1}\right)
\det\Bchi\,\mathrm{d}V_0\\
&\phantom{=}
+ \int \limits_{{\Omega_0}}\left(
\frac{\partial\psi}{\partial \Bkappa}
-\bu\cdot\frac{\partial\bf}{\partial \Bkappa}
\right)\cdot\bW\det\Bchi\,\mathrm{d}V_0 
 -  \int \limits_{\partial {\Omega}^\mathrm{S}_{T_0}} \bT^0\cdot\bw \, \mathrm{d}S_0
\end{aligned}
\end{align}
Following a number of previous authors \citep{Maugin1995,KienzlerHerrmann1997,Gurtin2000} we recognize the Eshelby stress tensor $\Eshelby:= \psi\iso - \bF^T\bP$.
We now rewrite this weak form on $\Omega$:
\begin{align}
\begin{aligned}[b]
0 &= \int \limits_{{\Omega}}
\bP: \frac{\partial \bw}{\partial \bX}
\,\mathrm{d}V
 - \int \limits_{{\Omega}} (\bf\cdot\bw) \, \mathrm{d}V\\
&\phantom{=}
+ \int \limits_{{\Omega}}
\left(\bP - \left(\bf\cdot\bu\right)\iso + \Eshelby \right):
\frac{\partial \bW}{\partial \bX}
\,\mathrm{d}V\\
&\phantom{=}
+ \int \limits_{{\Omega}}\left(
\frac{\partial\psi}{\partial \bX}
-\bu\cdot\frac{\partial \bf}{\partial \bX}
\right)\cdot\bW\,\mathrm{d}V
 -  \int \limits_{{\partial {\Omega}^\mathrm{S}_{T}}} \bT\cdot\bw \, \mathrm{d}S
\label{weakform}
\end{aligned}
\end{align}
The strong form is derived from the weak form using standard variational arguments. In equation (\ref{weakform}) we allow discontinuities at the phase interface in all fields except $\bW$ and $\bw$. The corresponding strong form consists of the two following sets of equations. The first set (\ref{eqn:standBC} - \ref{eqn:standPDE}) represents the standard governing partial differential equations and boundary conditions of nonlinear elasticity, now extended to include an interface $\Gamma$. The second set of equations (\ref{eqn:confBC} - \ref{eqn:confPDE}) has been simplified under the assumption that the preceding equations of nonlinear elasticity are satisfied. We use the operator $\nabla\cdot$ here to refer to the divergence with respect to $\bX$. Note that $\bN$ is the unit normal to the boundary of the body, and $\bN^\Gamma$ is the normal to the interface.
\begin{subequations}
\begin{align}
\bP\bN - \bT &=0 \text{ on } \partial\Omega^\mathrm{S}_{T}
\label{eqn:standBC}\\
\ljmp \bP \bN^\Gamma \rjmp &= 0 \text{ on } \Gamma\\
\nabla\cdot \bP + \bf &= 0 \text{ in } \Omega
\label{eqn:standPDE}\\
\left(\Eshelby + \bP - (\bf\cdot\bu)\iso\right)\bN &=0 \text{ on } \partial\Omega^\mathrm{M}_{T}
\label{eqn:confBC}\\
\ljmp \left(\Eshelby - (\bf\cdot\bu)\iso\right)\bN^\Gamma \rjmp &= 0 \text{ on } \Gamma \label{eqn:confJump}\\
\nabla\cdot \Eshelby 
-\frac{\partial\psi}{\partial \bX} 
- \bF^T\bf &= 0 \text{ in } \Omega
\label{eqn:confPDE}
\end{align}
\end{subequations}
Note that equation (\ref{eqn:confPDE}) corresponds exactly to the partial differential equation derived in \citet{Gurtin2000} and \citet{Maugin2011}. This reduces further,  using coordinate notation for clarity, with lower case indices for objects defined on $\Omega_t$ and upper case indices for objects defined on $\Omega$.

\begin{align}
 \left(\psi\delta_{IJ} - F_{iI}P_{iJ} \right)_{,J}
-\frac{\partial\psi}{\partial X_I} - F_{iI}f_i &= 0 \nonumber \\
\frac{\partial \psi}{\partial F_{iJ}}F_{iJ,I}
+\frac{\partial \psi}{\partial X_I}
 - F_{iI,J}P_{iJ} - F_{iI}P_{iJ,J}
-\frac{\partial\psi}{\partial X_I} - F_{iI}f_i &= 0 \nonumber \\
 P_{iJ}F_{iJ,I}
 - F_{iI,J}P_{iJ} - F_{iI}P_{iJ,J} - F_{iI}f_i &= 0
\end{align}
Using the relation $F_{iI,J} = F_{iJ,I}$ and $P_{iJ,J}= - f_i$ from equation (\ref{eqn:standPDE}), this equation vanishes identically. Thus, if the standard governing partial differential equation for nonlinear elasticity, $P_{iJ,J} + f_i = 0$ in $\Omega$ is satisfied, the corresponding configurational partial differential equation $\mathcal{E}_{IJ,J} -\frac{\partial\psi}{\partial X_I}
 - F_{iI}f_i = 0$ in $\Omega$ is trivially satisfied.

\subsection{Interfacial energy}
\label{sec:interfaceEnergy}
Interfacial energy can be included by adding the integral
\begin{align}
\Pi^\Gamma &= \int_\Gamma \psi^\Gamma\,\mathrm{d}S
\end{align}
to the free energy functional $\Pi$, where $\psi^\Gamma$ is the interfacial free energy density. Assuming $\psi^\Gamma$ is a constant, the first variation of this integral is
\begin{align}
\frac{\delta \Pi^\Gamma}{\delta \bU}\cdot \bW &=
 \int_{\Gamma} -2\psi^\Gamma H
\left(
 \bW
 \cdot\bN^\Gamma
\right) \,\mathrm{d}S\nonumber\\
&\phantom{=}
-\oint_{\partial \Gamma}
\psi^\Gamma \bW\cdot\left(\bN\times\bT^\Gamma\right) (\bN^\Gamma\cdot\bN)
\bT^\Gamma\cdot\mathrm{d}\br
 \end{align}
 where $H$ is the mean curvature of $\Gamma$. This term would affect equation (\ref{eqn:confJump}) and add a condition over $\partial\Gamma$, resulting in the following:
\begin{subequations}
\begin{align}
\left(\ljmp \Eshelby - (\bf\cdot\bu)\iso \rjmp -  2\psi^\Gamma H\iso\right)\bN^\Gamma &= 0 \text{ on } \Gamma\label{eqn:elastpluscurvdom}\\
\psi^\Gamma \bN^\Gamma\cdot\bN &= 0 \text{ on } \partial\Gamma_T\label{eqn:elastpluscurvboun}
\end{align}
\end{subequations}
A full derivation is included in \ref{sec:appc}.


\subsection{Nonequilibrium with respect to material evolution}

Suppose that the body is at equilibrium everywhere except with respect to configurational evolution of the interface, $\Gamma$. Then, as the foregoing treatment demonstrates, the first variation of the free energy reduces to 
\begin{align}
\frac{\delta \Pi}{\delta \bU}\cdot \bW 
&= \int \limits_{\Gamma}\bW\cdot \left[
\left(\ljmp \Eshelby - (\bf\cdot\bu)\iso \rjmp -  2\psi^\Gamma H\iso\right)\bN^\Gamma
\right] \,\mathrm{d}S
\end{align}
The field $\bW$ represents variation of the interface. In a rate formulation, this would be replaced by the interface velocity $ \bV_\Gamma$.
\begin{align}
\dot{\Pi}
&= \int \limits_{\Gamma}\bV_\Gamma\cdot \left[
\left(\ljmp \Eshelby - (\bf\cdot\bu)\iso \rjmp -  2\psi^\Gamma H\iso\right)\bN^\Gamma
\right] \,\mathrm{d}S
\end{align}
We let
\begin{align}
\bV_\Gamma = -\bM \left[
\left(\ljmp \Eshelby - (\bf\cdot\bu)\iso \rjmp -  2\psi^\Gamma H\iso\right)\bN^\Gamma
\right]
\label{eqn:lsVelocity}
\end{align}
with $\bM$ a positive definite tensor to ensure decrease in free energy ($\dot{\Pi} \leq 0$),  thus satisfying the second law of thermodynamics.
The presence of $H$ in the interface velocity makes this a general curvature driven flow. 

\subsection{Numerical treatment}
We use the level set method for movement of the sharp interface in the case of nonequilibrium with respect to material evolution. All partial differential equations are solved using the finite element method.

\subsubsection{Level set method}

In the level set method, the interface is represented by the zero contour or level set of a scalar field, $\Phi(\bX,t)$ \citep{OsherSethian1988}. The evolution of $\Phi$ (and the zero level set) is governed by the following partial differential equation:
\begin{align}
\frac{\partial \Phi}{\partial t} = -v|\Bnabla \Phi|
\end{align}
where the scalar $v$ is the advection velocity (i.e. the normal component of the level set velocity). The behavior of the level set evolution is improved when $\Phi$ is a signed distance function. Replacing $v$ with $v_e$ helps to maintain this property, where $v_e$ is the \textit{extensional velocity}, defined as the advection velocity at the closest point on the zero level set. This ``closest point'' generally does not coincide with a node or an integration point. Additionally, the field $\Phi$ is periodically reinitialized as a signed distance function based on the current location of the zero level set. The method of reinitialization used here involves solving the Eikonal equation $|\Bnabla \Phi| = 1$, using the following partial differential equation  \citep{RussoSmereka2000} with additional constraints on $\Phi$ to reduce spurious movement of the zero level set:

\begin{align}
\frac{\partial \Phi}{\partial \hat{t}} &= \text{sgn}(\Phi_0)(1 - |\Bnabla \Phi|) \nonumber \\
\Phi_0(\bX) &= \Phi(\bX,0)
\end{align}
Note that $\hat{t}$ is a time-like parameter introduced only to allow relaxation of $\Phi$ to a signed distance function during reinitialization.

\subsubsection{Finite element methods}

Both the level set equation and the Eikonal equation used in reinitialization are solved using finite element methods. To reduce spatial oscillations common to advection-diffusion equations, the streamline upwind/Petrov-Galerkin (SUPG) weak form is used \citep{BrooksHughes1982}:
\begin{align}
\tilde{w} &= w + \tau \frac{\bv\cdot\nabla w}{|\bv|} \nonumber \\
\int \limits_\Omega \tilde{w}\frac{\partial \Phi}{\partial t} \,\mathrm{d}V
&= -\int \limits_\Omega  \tilde{w}v_e|\Bnabla \Phi| \,\mathrm{d}V
\end{align}
Since we are only interested in $\Phi$ near the zero level set, the level set equation is only solved within a narrow band about the zero level set. The elastic, finite deformation of the body is modeled using the Bubnov-Galerkin weak form. The elasticity problem is solved over the entire domain using the field $\Phi$ to determine material properties at each integration point.\\

\subsection{Numerical simulation}

      \begin{figure}[htb]
        \centering
        \includegraphics[width=0.6\textwidth]{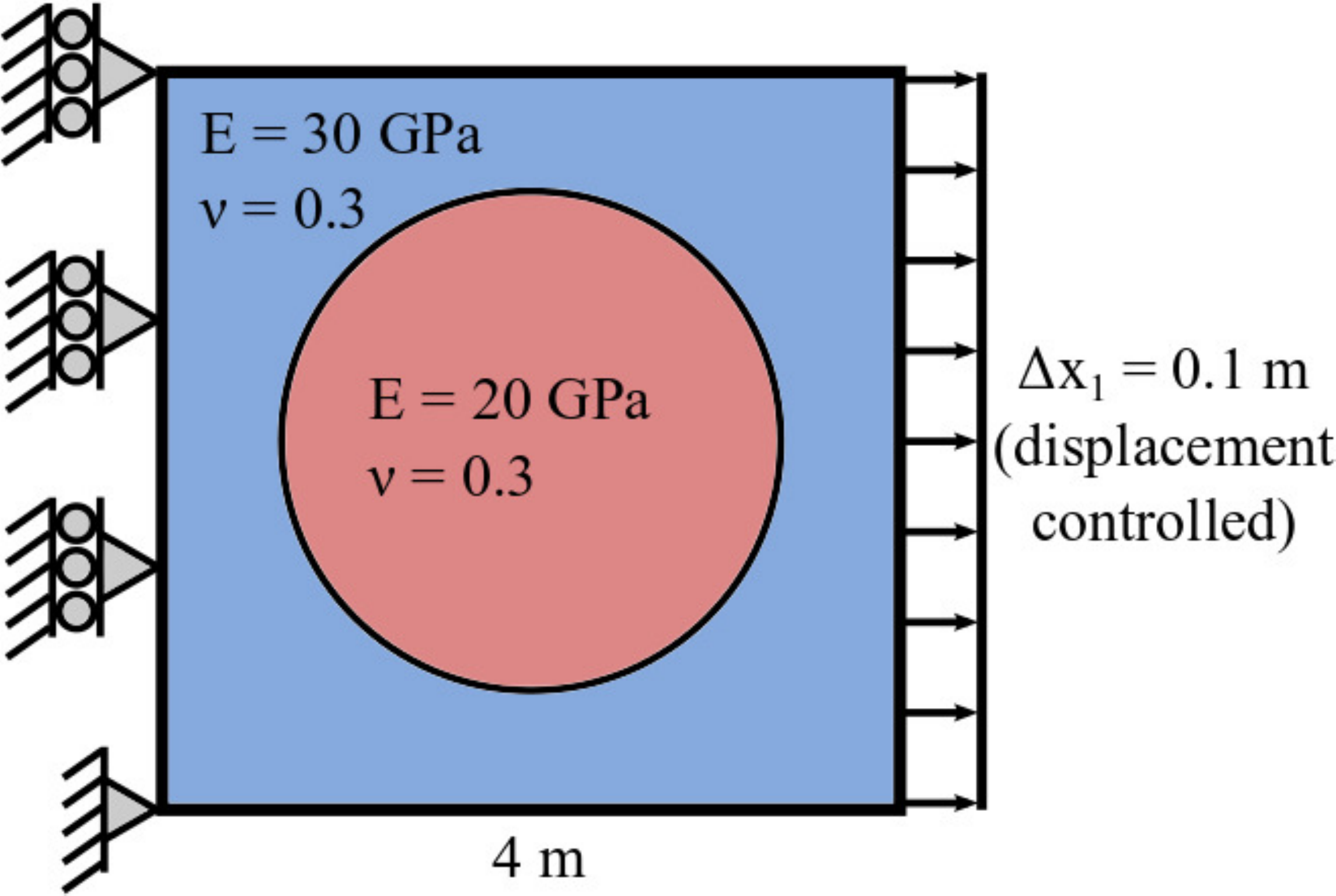}
        \caption{An example problem of sharp interface motion driven by displacement controlled, uniaxial tension.}
	\label{fig:sharpExample}
      \end{figure}

\begin{figure}[htb]
        \centering
\begin{minipage}[t]{0.3\textwidth}
        \centering
	\includegraphics[width=0.9\textwidth]{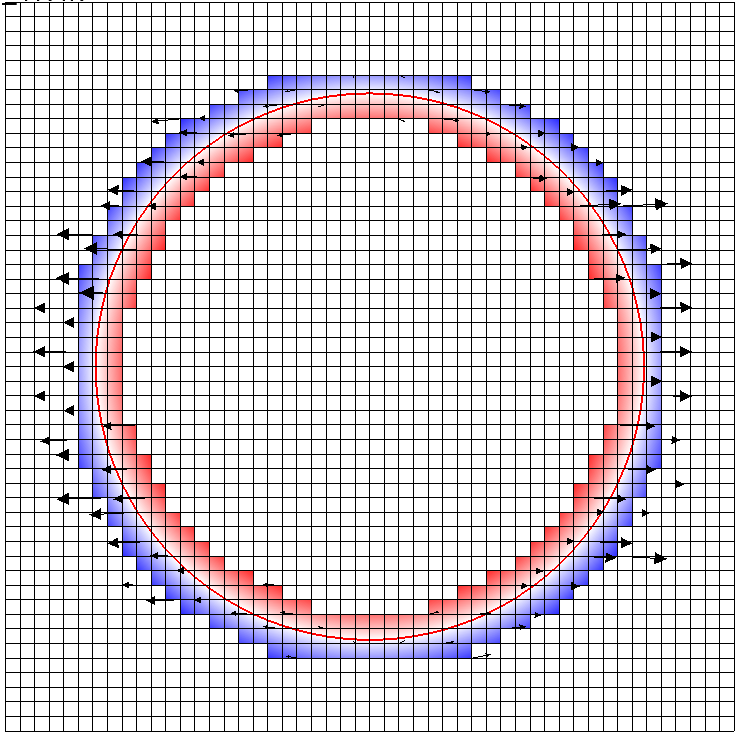}
	\captionof{subfigure}{Time step = 1}
\end{minipage}
\begin{minipage}[t]{0.3\textwidth}
        \centering
	\includegraphics[width=0.9\textwidth]{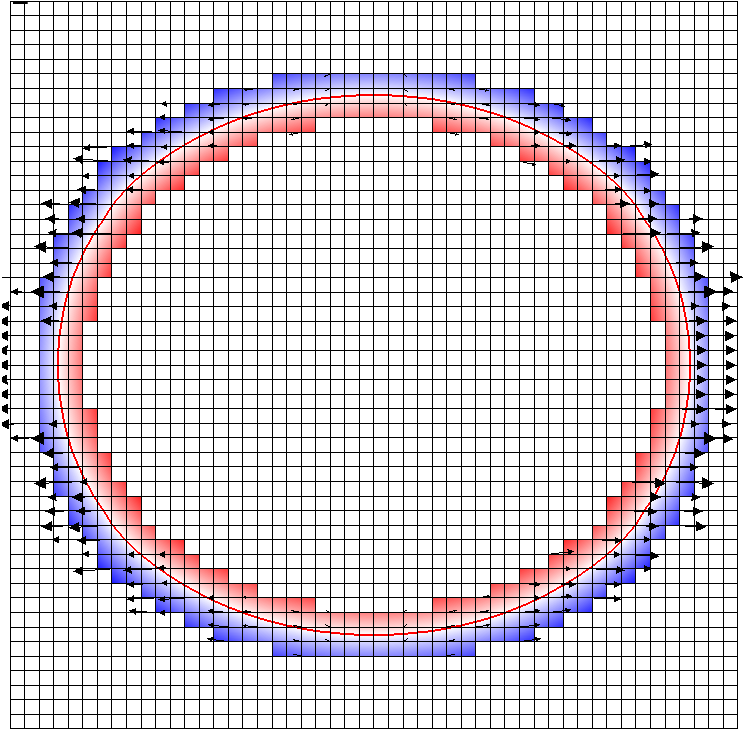}
	\captionof{subfigure}{Time step = 50}
\end{minipage}
\begin{minipage}[t]{0.3\textwidth}
        \centering
	\includegraphics[width=0.9\textwidth]{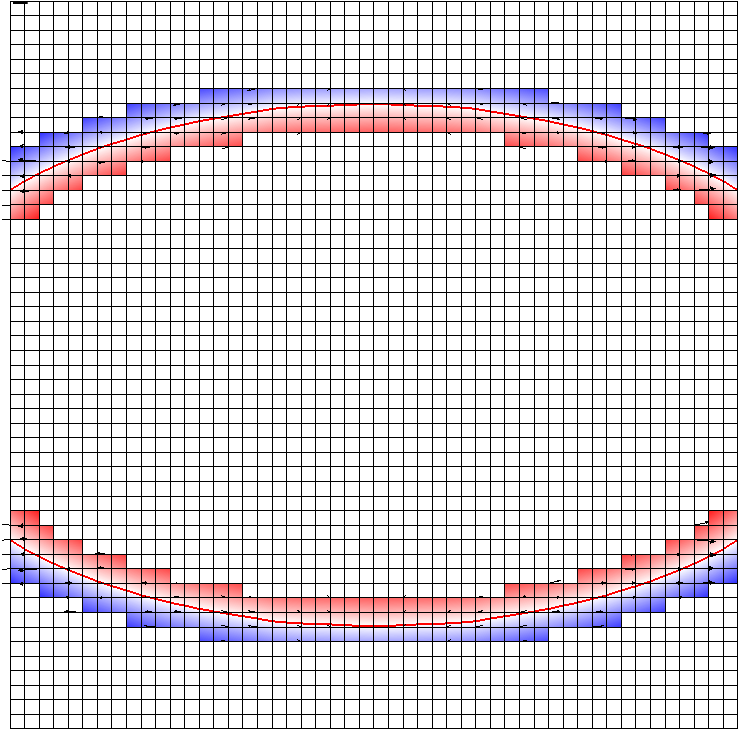}
	\captionof{subfigure}{Time step = 250}
\end{minipage}
\begin{minipage}[t]{0.3\textwidth}
        \centering
	\includegraphics[width=0.9\textwidth]{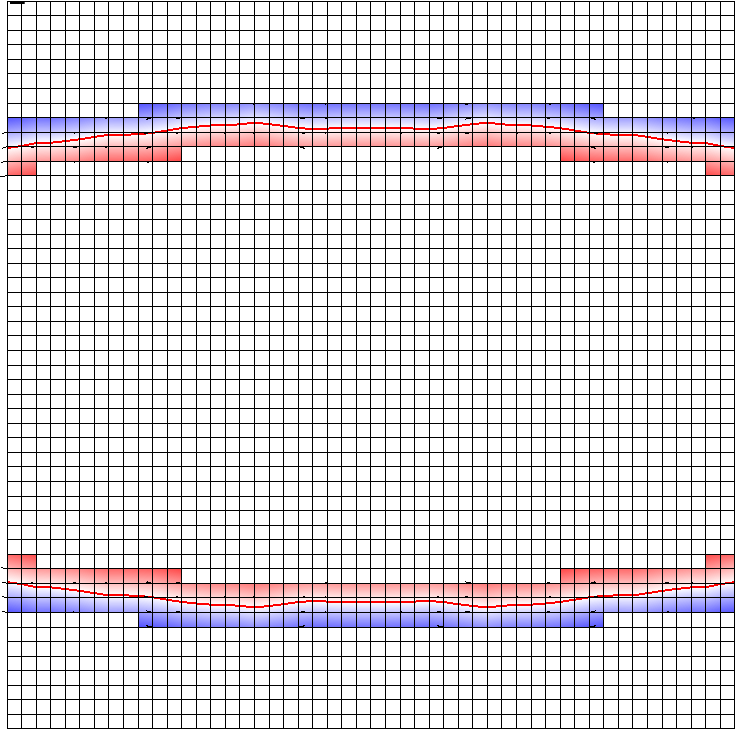}
	\captionof{subfigure}{Time step = 1000}
\end{minipage}
\begin{minipage}[t]{0.3\textwidth}
        \centering
	\includegraphics[width=0.9\textwidth]{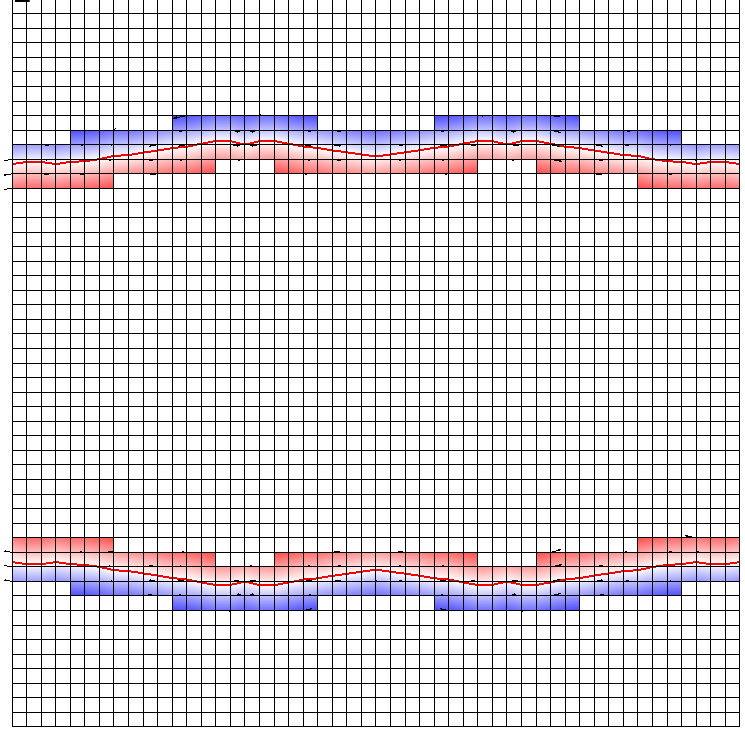}
	\captionof{subfigure}{Time step = 4000}
\end{minipage}
\begin{minipage}[t]{0.3\textwidth}
        \centering
	\includegraphics[width=0.9\textwidth]{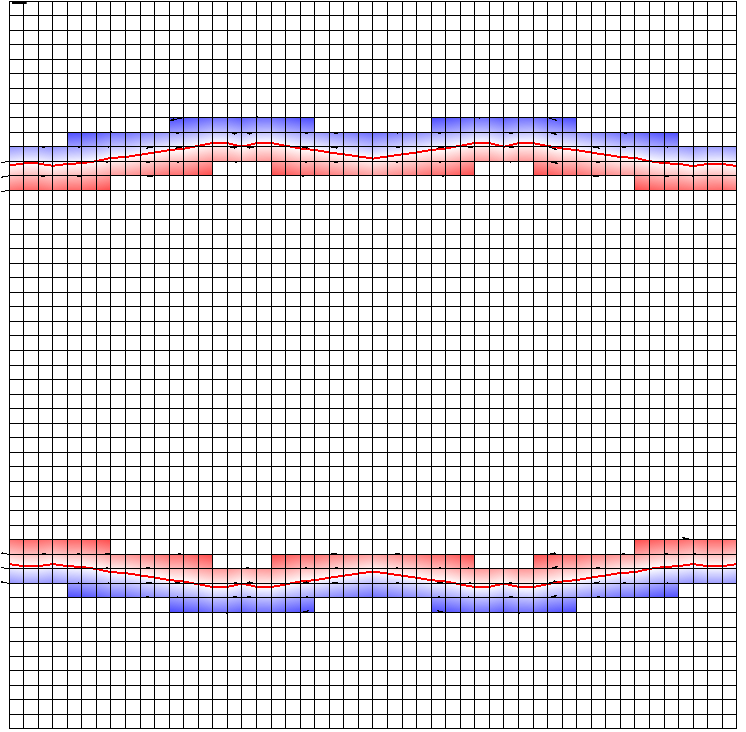}
	\captionof{subfigure}{Time step = 7000}
\end{minipage}
        \caption{Evolution of the sharp interface in the evolved reference configuration $\Omega$ over time. The interface is represented by the zero level set (red line) and is updated over a narrow band enclosing the zero level set (colored elements). The velocity vectors of the level sets are shown, but nearly vanish at later times, and are therefore not discernible in (c--f).}
	\label{fig:lsEvolution}
\end{figure}

\begin{figure}[htb]
        \centering
        \includegraphics[width=0.5\columnwidth]{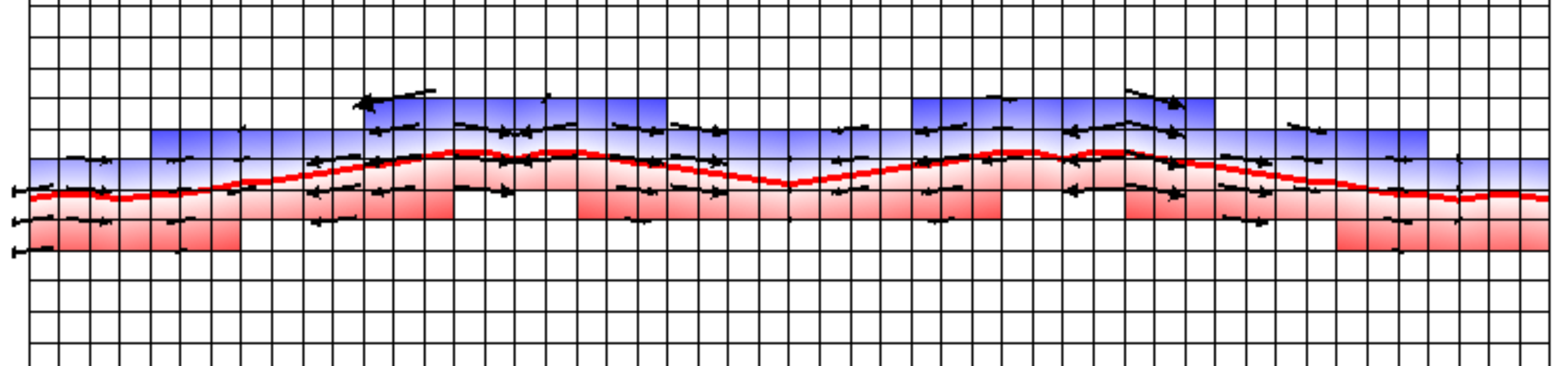}
        \caption{At steady state, the level set velocity vectors are all tangential to the level sets (at time step = 7000 and velocity vectors scaled 3x).}
	\label{fig:tangentVel}
      \end{figure}

\begin{figure}
        \centering
\begin{minipage}[t]{0.4\textwidth}
        \centering
	\includegraphics[width=0.9\textwidth]{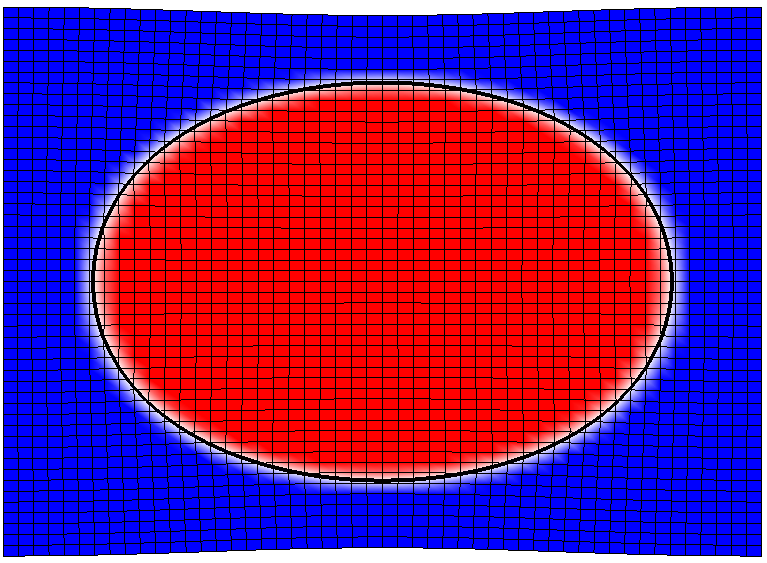}
	\captionof{subfigure}{Time step = 1}
\end{minipage}
\begin{minipage}[t]{0.4\textwidth}
        \centering
	\includegraphics[width=0.9\textwidth]{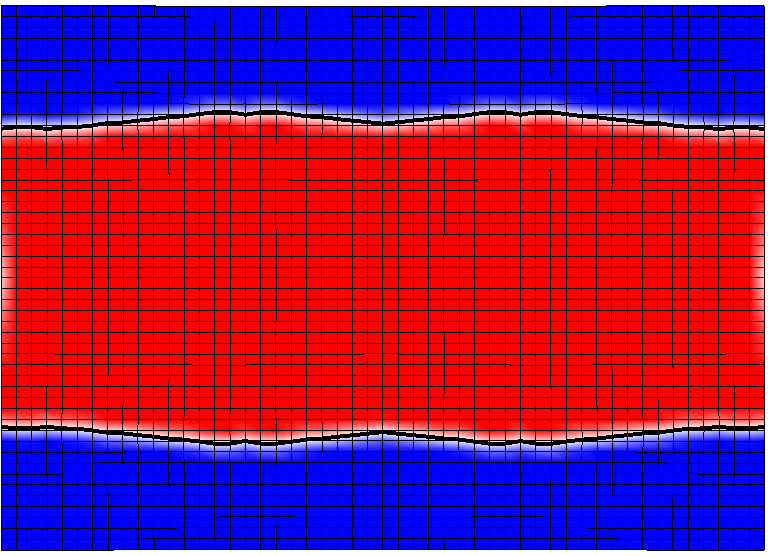}
	\captionof{subfigure}{Time step = 7000}
\end{minipage}
        \caption{Evolution of elastic deformation due to the change in material configuration (10x displacement shown).}
	\label{fig:sharpDisplace}
\end{figure}

Figures \ref{fig:sharpExample} - \ref{fig:sharpDisplace} present the plane strain computation of a two-phase material with a migrating sharp interface, under vanishing body force and zero interfacial energy. The initial phase distribution is as shown in Figure \ref{fig:sharpExample}, with a compliant phase (Young's modulus $E = 20$ GPa) surrounded by a stiffer phase ($E = 30$ GPa). All other material properties are the same for both phases. The body is subjected to displacement controlled, uniaxial tension. At each time step, the interface motion is modeled via the level set equation, and the current elastic deformation is found based on the updated interface location. The zero level set velocity is found using equation \ref{eqn:lsVelocity}. We used a time step of .001 s and $M$ equal to the isotropic tensor multiplied by 2e-8 $\mathrm{m}^3/(\mathrm{Ns})$. The problem was allowed to evolve until an apparent steady state was achieved.

Figure \ref{fig:lsEvolution} shows the evolution of the zero level set as the interface moves in the evolved reference configuration $\Omega$, driven by the jump in the Eshelby stress tensor. A steady state solution is arrived at within 4000 time steps, or 4 s. The time scale of the probem is consistent with the interface velocity, which is initially about 5 m/s, and the size of the domain. It is shown in Figure \ref{fig:tangentVel} that at steady state, all of the level set velocity vectors are tangential to the level set. As a result, there is no movement normal to the zero level set and a steady state is achieved. 

The elastic deformation on the top and bottom edges of the body can be seen in Figure \ref{fig:sharpDisplace}, which is in the current, deformed configuration $\Omega_t$. Initially, that deformation is greatest in the middle, due to the concentration of the compliant phase in the center. However, the interface evolves in such a way that the elastic deformation along the edges becomes nearly uniform by the time the phases reach a steady state.

\FloatBarrier

\section{Configurational change over a volume; diffuse interfaces}
\label{sec:diffuseinterface}

      \begin{figure}[htb]
        \centering
        \includegraphics[width=\textwidth]{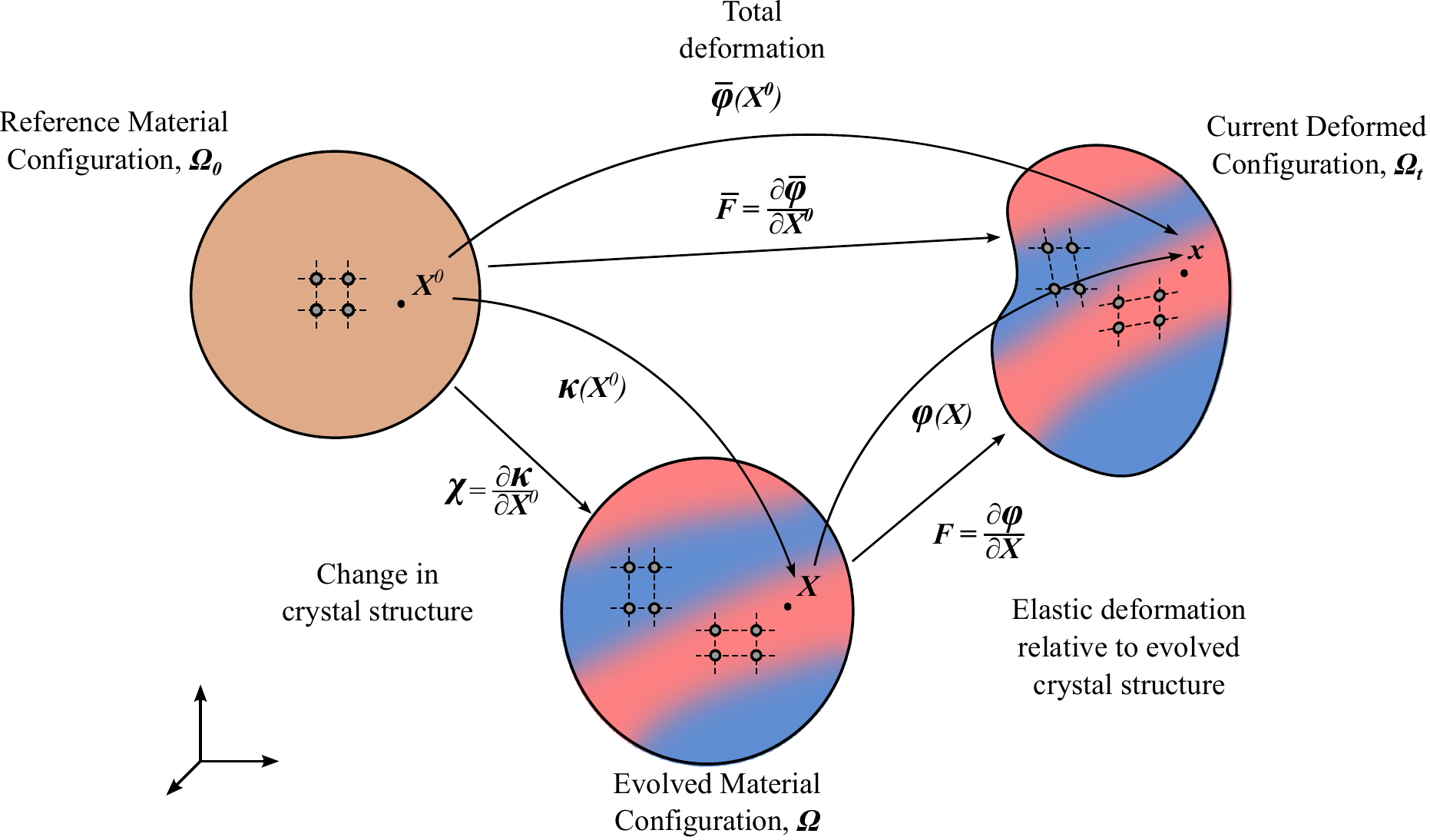}
        \caption{Kinematics of the configurational changes and elastic deformations causing a diffuse interface.}
	\label{fig:diffuse}
      \end{figure}

We now turn to the case of a solid that can undergo phase changes throughout the volume, considering again equilibrium with respect to both the configurational and standard displacements. As discussed in the Introduction, the material's configuration, represented by the crystal structure, varies smoothly between parent and daughter phases, thus creating diffuse interfaces between the phases (see Figure~\ref{fig:diffuse}). In this case, the invertible map $\Bkappa$  and its gradient, $\Bchi = \partial \Bkappa/\partial \bX^0$, model the distortion of the crystal structure associated with the phase change.
We introduce the mapping $\bar{\Bphi}$ and gradient $\bar{\bF}$ giving the final placement of the body via configurational and standard displacements from the reference material configuration:
\begin{align}
\bx &= \bar{\Bphi}(\bX^0) =  \bX^0 + \bar{\bu}\\
\bar{\bF} &= \frac{\partial \bar{\Bphi}}{\partial \bX^0} = \iso +  \frac{\partial \bar{\bu}}{\partial \bX^0}
\end{align}
The deformation map $\Bphi$ and deformation gradient $\bF$ model standard elastic deformation relative to the evolved material configuration (distorted crystal structure) and are as defined previously.

\subsection{The cubic to tetragonal transformation}
\label{sec:cubictet}
\begin{figure}[htb]
        \centering
\begin{minipage}[t]{0.4\textwidth}
        \centering
	\includegraphics[width=0.9\textwidth]{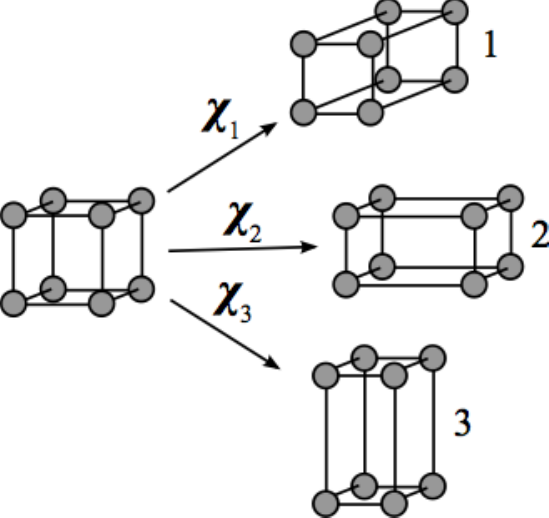}
\end{minipage}
\begin{minipage}[t]{0.4\textwidth}
        \centering
	\includegraphics[width=0.9\textwidth]{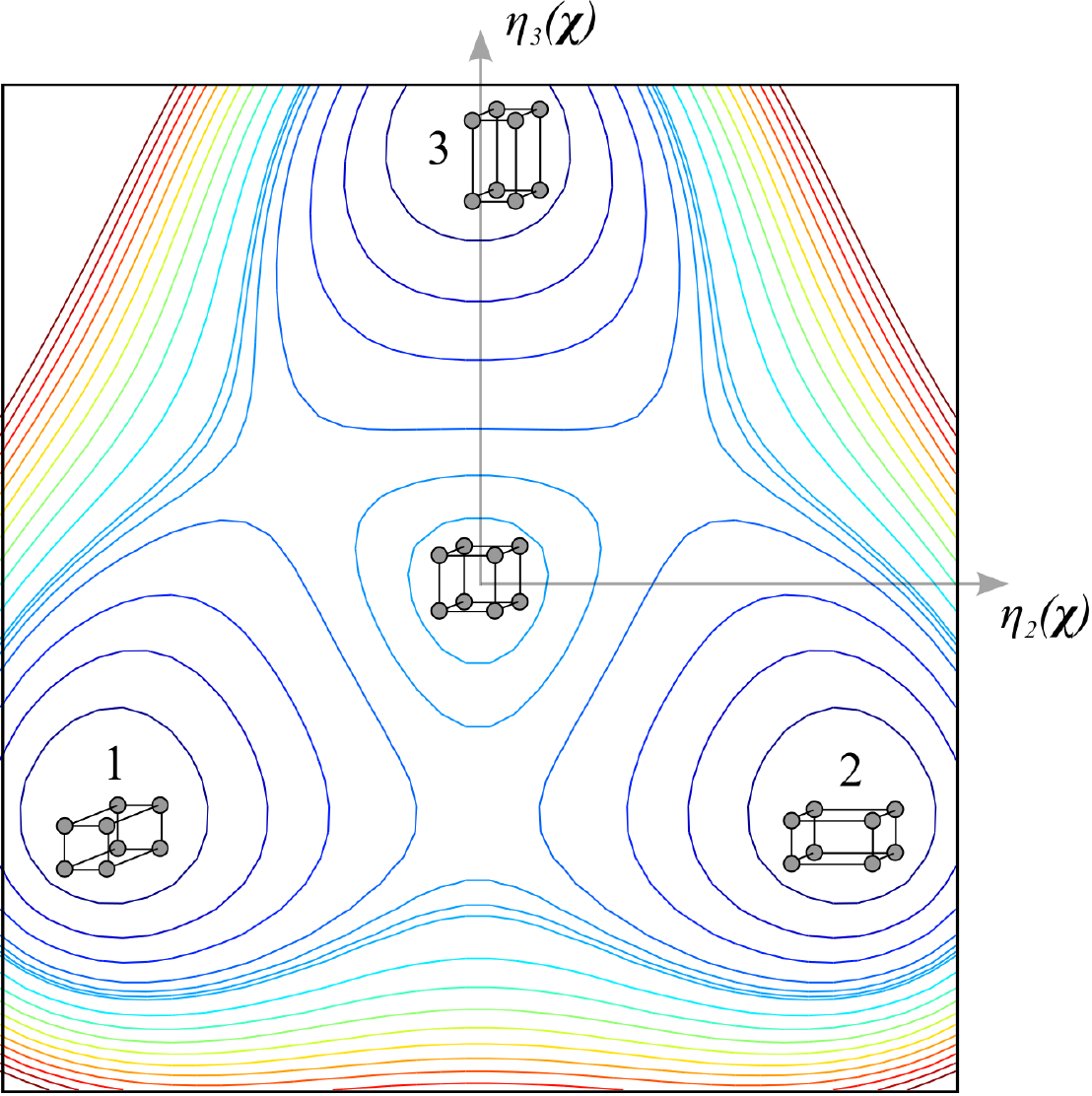}
\end{minipage}
        \caption{Tetragonal variants and free energy density contours in 3D. The axis, $\eta_2$ and $\eta_3$, are reparametrized strains.}
	\label{fig:tet3d}
\end{figure}

\begin{figure}[htb]
        \centering
        \includegraphics[width=0.7\columnwidth]{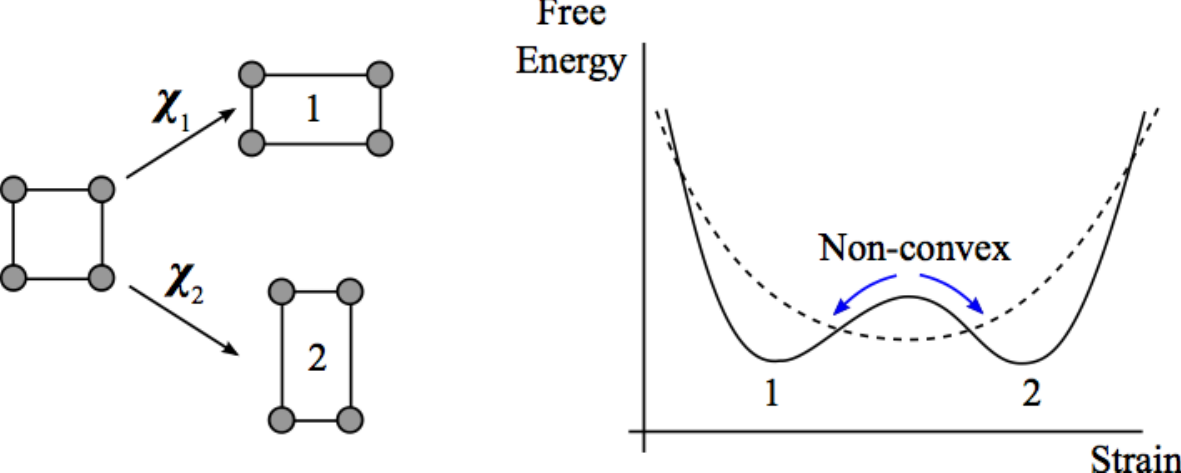}
	\caption{Tetragonal variants and free energy density schematic in 2D.}
	\label{fig:tet2d}
      \end{figure}

To fix ideas we consider a cubic to tetragonal transformation, although our methods have wider applicability to any smooth change in crystal structure. In the reference material configuration, the solid is stable in the cubic crystal structure at high temperature. The free energy density function has a single well in this phase. We assume that a rapid quench renders the cubic phase unstable to a configurational change by distortion into tetragonal phase. Three such tetragonal variants are possible, and the solid is stable in any of these structures, implying that they are local minima of the free energy density. Furthermore the configurational change between one tetragonal variant and another is smooth, and these variants are separated by diffuse interfaces. In this phase, therefore, the solid's free energy density is a smooth, non-convex  function of $\Bchi$ with three minima corresponding to the three stable and equivalent tetragonal variants (Figure \ref{fig:tet3d}). A two-dimensional version of these configurational changes is the square to rectangle transformation (Figure \ref{fig:tet2d}), which serves well to fix ideas. We also return to  two dimensions for the first numerical example in Section \ref{sec:numsims}. While we use free energy functions with wells of equal depth as examples, there is nothing in this work that requires that the depths be equal.

\subsection{Free energy density functions}
\label{sec:freeenergydens}
\begin{figure}[htb]
        \centering
        \includegraphics[width=\textwidth]{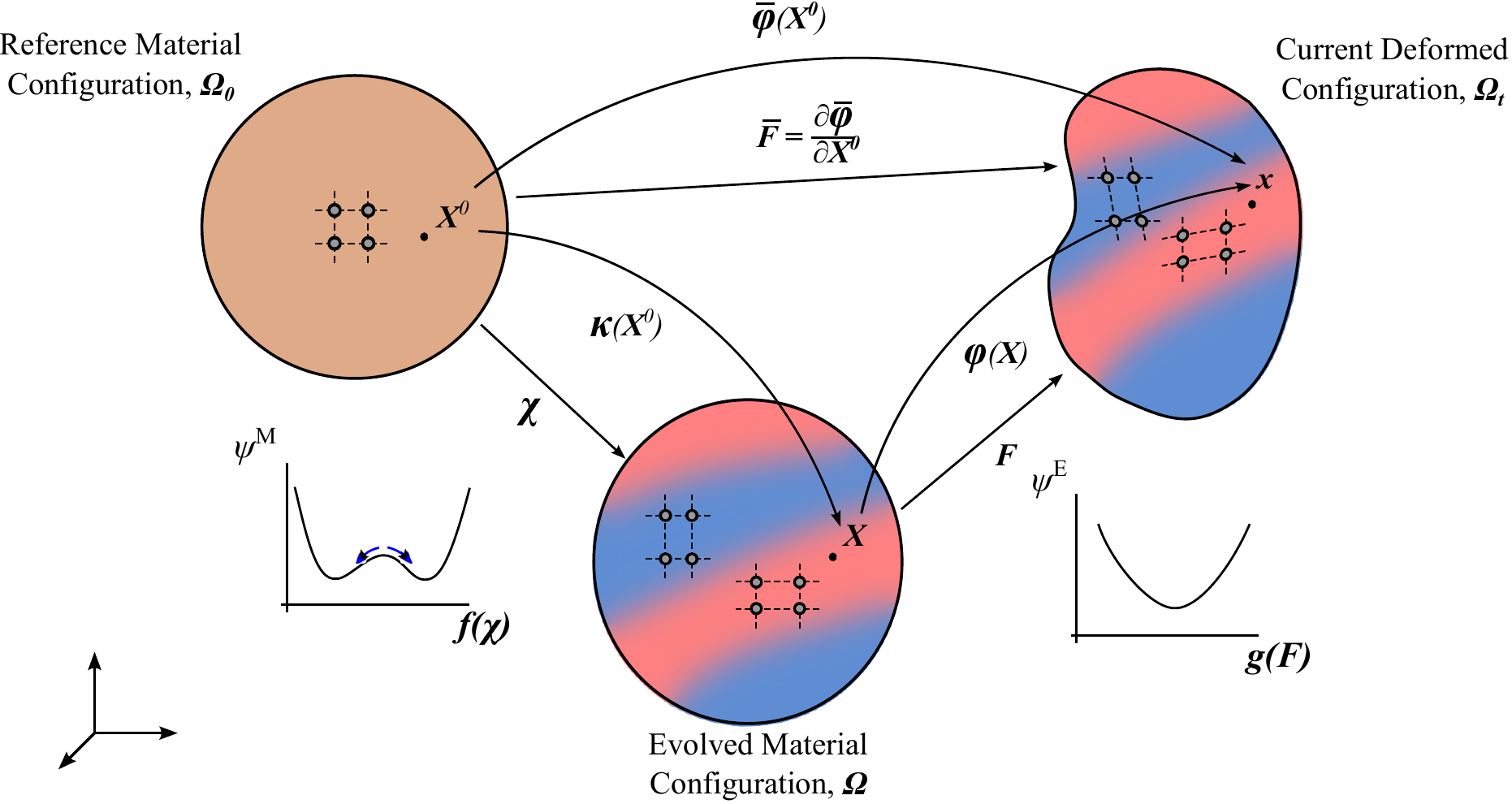}
	\caption{Schematic of the kinematics and free energy associated with evolution of the configuration and elastic deformation.}
	\label{fig:energySchem}
      \end{figure}

The free energy density function associated with the configurational changes described above is $\psi^\mathrm{M} = \hat{\psi}^\mathrm{M}(\bX^0,\Bchi,\nabla^0\Bchi)$. Note that we use the operator $\nabla^0$ to refer to derivatives with respective to $\bX^0$ for brevity in notation. This function allows for inhomogeneity via $\bX^0$ and, as discussed in Section \ref{sec:cubictet}, is dependent on the tangent map of the configurational field $\Bchi$. Since $\psi^\mathrm{M}$ is non-convex in the tetragonal phase, it allows the development of microstructures formed by laminae of the tetragonal variants, as discussed extensively in the literature. See \citet{BhattacharyaKohn1997,Muller1999,Bhattacharyaetal2004} for a background discussion. As is also well-known, these microstructures can develop with arbitrary fineness unless the diffuse interfaces between sub-regions of a single variant are penalized. This is done by inclusion of a dependence on $\nabla^0\Bchi$ for regularization \citep{BallCrooks2011,Rudrarajuetal2014}. This ensures physically meaningful solutions and mathematical well-posedness.
The free energy density function for the standard elastic deformation relative to $\Omega$ is $\psi^\mathrm{S} = {\psi}^\mathrm{S}(\bX,\bF,\Bchi)$, where $\bX = \Bkappa(\bX^0)$. Similar to $\psi^\mathrm{M}$, the elastic free energy depends on $\bX$ and the elastic deformation gradient $\bF$. Anisotropic elastic response can be incorporated if $\psi^\mathrm{S}$ is made to depend on $\Bchi$: The local value of $\Bchi$ determines the tetragonal variant arising as a result of the configurational change, and therefore sets the  anisotropy of response due to elastic deformation relative to this evolved material configuration, $\Omega$. The free energy of the system is then modeled with the following Gibbs free energy functional (Figure \ref{fig:energySchem}):
\begin{align}
\begin{aligned}[b]
\Pi[\bar{\bu};\bU] &=
\int \limits_{{\Omega}_0} \big[ \psi^\mathrm{M}(\bX^0,\Bchi,\nabla^0\Bchi) + \psi^\mathrm{S}(\bX,\bF,\Bchi)\det\Bchi \big] \,\mathrm{d}V_0\\
&\phantom{=}
  - \int \limits_{{\Omega}_0} \bf^0\cdot \bar{\bu} \, \mathrm{d}V_0
 -  \int \limits_{\partial \Omega^\mathrm{S}_{T_0}} \bT^0\cdot\bar{\bu} \, \mathrm{d}S_0
\end{aligned}
\label{eqn:diffuseGibbs}
\end{align}
We draw attention to the definition of quantities relative to the reference material configuration, $\Omega_0$, extending to the work terms of the body force and traction. This seems natural because $\Bkappa$ corresponds to distortion of the crystal structure, and $\Bphi$ is further motion relative to the distorted crystal. Therefore, the distributed forces are dual to the total displacement $\bar{\bu} = \bU + \bu$.
\subsection{Variational formulation}
We again seek equilibrium by setting the first variation of the Gibbs free energy to vanish. We consider variations on the configurational displacement, $\bU^\eps := \bU + \eps \bW$, and the total displacement, $\bar{\bu}^\eps := \bar{\bu} + \eps \bar{\bw}$. Then, equilibrium requires
\begin{align}
\begin{aligned}[b]
\frac{\mathrm{d}}{\mathrm{d}\eps} \Pi[\bar{\bu}^\eps;\bU^\eps] \evat_{\eps=0} &=
\frac{\mathrm{d}}{\mathrm{d}\eps} \bigg\{ 
\int \limits_{{\Omega}_0} \psi^\mathrm{M}(\bX^0,\Bchi^\eps,\nabla^0\Bchi^\eps) \,\mathrm{d}V_0\\
&\phantom{=}
+ 
\int \limits_{{\Omega}_0} \psi^\mathrm{S}(\bX^\eps,\bF^\eps,\Bchi^\eps)\det\Bchi^\eps\,\mathrm{d}V_0\\
&\phantom{=}
- 
\int \limits_{{\Omega}_0} \bf^0\cdot \bar{\bu}^\eps \, \mathrm{d}V_0 -  
\int \limits_{\partial \Omega^\mathrm{S}_{T_0}} \bT^0 \cdot \bar{\bu}^\eps \, \mathrm{d}S_0 \bigg\} \evat_{\eps=0}\\
&=0
\end{aligned}
\end{align}

We apply the earlier results concerning the first variations of $\bF$ and $\det\Bchi$. We also define $\bD := \partial\psi^\mathrm{M}/\partial\Bchi$, 
$\bB :=  \partial \psi^\mathrm{M}/\partial \nabla^0\Bchi$
 and $J_\chi := \det\Bchi$. 

 The resulting weak form is the following: {
\begin{align}
\begin{aligned}[b]
0 &=
\int \limits_{{\Omega}_0}
\bB\vdots\nabla^0\nabla^0\bW \, \mathrm{d}V_0
+
\int \limits_{{\Omega}_0}
J_\chi\frac{\partial \psi^\mathrm{S}}{\partial\bX}\cdot \bW \, \mathrm{d}V_0\\
&\phantom{=}
+\int \limits_{{\Omega}_0}
 \left[\bD+ 
J_\chi\left(\Eshelby \Bchi^{-T}
 + \frac{\partial \psi^\mathrm{S}}{\partial \Bchi} \right) \right]:
 \nabla^0\bW \, \mathrm{d}V_0\\
&\phantom{=}
+
\int \limits_{{\Omega}_0}
J_\chi \left( \bP \Bchi^{-T} \right):
 \nabla^0\bar{\bw} \, \mathrm{d}V_0
- 
\int \limits_{{\Omega}_0}
 \bf^0\cdot  \bar{\bw} \, \mathrm{d}V_0
 -  
\int \limits_{\partial \Omega^\mathrm{S}_{T_0}} \bT^0\cdot \bar{\bw} \, \mathrm{d}S_0
\end{aligned}
\end{align}  

Here, $\bD$ is a configurational stress that is distinct from the Eshelby stress $\Eshelby$, and $\bB$ represents a higher order configurational stress. Deriving the strong form from this weak form involves several additional terms due to the dependence on $\nabla^0 \Bchi$, as described in \citet{Rudrarajuetal2014}.
We use the normal and surface gradient operators,  $\nabla^n$ and $\nabla^s$, where
\begin{align}
\nabla^n \psi &= \nabla^0\psi\cdot\bN^0\\
\nabla^s \psi &= \nabla^0 \psi - \left(\nabla^n\psi\right)\bN^0
\end{align}
Also, $\bb = -\nabla^s \bN^0 = \bb^T$
is the second fundamental form of the smooth parts of the boundary, $\partial\Omega_0$. We let $\bN^\mathcal{C} = \BXi\times\bN^0$, where $\BXi$ is the unit tangent to the smooth curve $\mathcal{C}_0$ that forms an edge between subsets $\partial \Omega^+_0$ and $\partial \Omega^-_0$of the smooth boundary surfaces $\partial \Omega_0$. If $\bN^{\mathcal{C}^+}$ is the outward unit normal to $\mathcal{C}_0$ from $\partial \Omega^+_0$ and $\bN^{\mathcal{C}^-}$ is the outward unit normal to $\mathcal{C}_0$ from $\partial \Omega^-_0$, then we define $\ljmp \bB:\left( \bN^\mathcal{C} \otimes \bN^0\right)  \rjmp^\mathcal{C} := \bB:\left( \bN^{\mathcal{C}^+} \otimes \bN^0\right) + \bB:\left( \bN^{\mathcal{C}^-} \otimes \bN^0\right)$. Applying the appropriate integration by parts and standard variational arguments leads to the following strong form.

\begin{subequations}
\begin{align}
J_{\chi} \bP \Bchi^{-T} \bN^0
- \bT^0  &= 0 \text{ on } \partial {\Omega^\mathrm{M}_{T_0}}\\
\nabla^0\cdot \left(J_{\chi} \bP \Bchi^{-T} \right)
+  \bf^0 &= 0 \text{ in } \Omega_0
\label{eqn:diffuse_pde}\\
\ljmp \bB:\left( \bN^\mathcal{C} \otimes \bN^0\right)  \rjmp^\mathcal{C}
&= 0  \text{ on }  \mathcal{C}^\mathrm{M}_{T_0}\\
\bB:\left( \bN^0 \otimes \bN^0\right)
&= 0  \text{ on } \partial {\Omega^\mathrm{S}_{T_0}}\\
\bD\bN^0 + 
J_{\chi}\left(\Eshelby\Bchi^{-T}
 + \frac{\partial \psi^\mathrm{S}}{\partial \Bchi} \right) \bN^0
-\bC &= 0  \text{ on } \partial {\Omega^\mathrm{S}_{T_0}}\label{configtracbc}\\
\nabla^0\cdot
 \left(\bD
 + 
J_{\chi}\frac{\partial \psi^\mathrm{S}}{\partial \Bchi} \right)
 +
\bF^T\bf^0 - \nabla^0\nabla^0\bB
&= 0 \text{ in } \Omega_0
\end{align}
\end{subequations}

where, using coordinate notation for clarity,
\begin{align}
\begin{aligned}[b]
C_I &= \nabla^nB_{I\gamma\zeta}N^0_\zeta N^0_\gamma
+2\nabla^s_\gamma B_{I\gamma\zeta} N^0_\zeta\\
&\phantom{=}
+B_{I\gamma\zeta}\nabla^s_\gamma N^0_\zeta
- (b_{\xi\xi} N^0_\gamma N^0_\zeta - b_{\gamma\zeta})B_{I\gamma\zeta}
\end{aligned}
\end{align}
Details of the above derivations of weak and strong forms appear in \ref{sec:appb}.

\subsection{Numerical simulations}
\label{sec:numsims}
We use the following double well, free energy density function to represent the two-dimensional, square to rectangle transformation:
\begin{align}
\BTheta &= \half(\Bchi^T\Bchi - \mathbbm{1})\\
\eta_1 &= \Theta_{11} + \Theta_{22},\,
\eta_2 = \Theta_{11} - \Theta_{22},\,
\eta_6 = \Theta_{12}\\
\psi^\mathrm{M} &= \frac{d}{s^2}\left(\eta_1^2 + \eta_6^2\right)
-\frac{2d}{s^2}\eta_2^2 + \frac{d}{s^4}\eta_2^4
+ \frac{l^2d}{s^2}|\nabla^0 \eta_2|^2
\label{eqn:psi_M}
\end{align}
where the energy wells lie at $\eta_2 = \pm s$ with a depth of $-d$. Additionally, we draw attention to the last term in Equation (\ref{eqn:psi_M}), which is the gradient free energy contribution that regularizes the non-convex free energy density as discussed in Section \ref{sec:freeenergydens}. Using standard dimensional arguments this term has been scaled by a length parameter $l$, where $1/l^2$ is the ratio of standard to strain gradient moduli.

For the three-dimensional case, we use the following reparameterized strain space:
\begin{align}
\begin{aligned}[b]
\eta_1 &= \frac{1}{\sqrt{3}}(\Theta_{11} + \Theta_{22} + \Theta{33}),\qquad
\eta_2 = \frac{1}{\sqrt{2}}(\Theta_{11} - \Theta_{22}),\\
\eta_3 &= \frac{1}{\sqrt{6}}(\Theta_{11} + \Theta_{22} - 2\Theta{33}),\qquad
\eta_4 = \sqrt{2}\Theta_{23} =\sqrt{2}\Theta_{32},\\
\eta_5 &= \sqrt{2}\Theta_{13} =\sqrt{2}\Theta_{31},\qquad
\eta_6 = \sqrt{2}\Theta_{12} =\sqrt{2}\Theta_{21}\\
\end{aligned}
\end{align}
The corresponding free energy density function has three wells located at $(\tfrac{\sqrt{3}}{2}s,\tfrac{1}{2}s)$, $(-\tfrac{\sqrt{3}}{2}s,\tfrac{1}{2}s)$, and $(0,s)$ in $(\eta_2,\eta_3)$ space with a depth of $-d$.
\begin{align}
\begin{aligned}[b]
\psi^\mathrm{M} &= \frac{3d}{2s^2}\left(\eta_1^2 +\eta_4^2 +\eta_5^2 + \eta_6^2\right)
-\frac{3d}{2s^2}(\eta_2^2 + \eta_3^2) + \frac{3d}{2s^4}\left(\eta_2^2+\eta_3^2\right)^2\\
&\phantom{=} + \frac{d}{s^3}\eta_3\left(\eta_3^2 - 3\eta_2^2\right)
+ \frac{3l^2d}{2s^2}\left(|\nabla^0 \eta_2|^2 + |\nabla^0 \eta_3|^2 \right)
\label{eqn:psi_M_3D}
\end{aligned}
\end{align}
Note the regularizing gradient free energy in the last two terms of Equation (\ref{eqn:psi_M_3D}). We also use an anisotropic St.~Venant-Kirchhoff model for the elastic deformation,
\begin{align}
\psi^\mathrm{S} &= \half \bE:\mathbb{C}(\Bchi):\bE
\label{eqn:psi_E}
\end{align}
where
\begin{align}
\begin{aligned}[b]
\mathbbm{C}(\Bchi) &= 
\sum_{I=1}^3 \alpha_I(\Bchi)\bM_I\otimes\bM_I
+\sum_{\substack{J,K=1\\J\neq K}}^3 \beta_{JK}(\Bchi)\bM_J\otimes\bM_K \\
&\phantom{=}
+2\mu(\mathbb{I} - \sum_{I=1}^3 \bM_I\otimes\bM_I)
\end{aligned}
\end{align}
with $\bE = \half(\bF^T\bF - \mathbbm{1})$, $\bM_I = \be_I\otimes \be_I$,
$\beta_{JK} = \beta_{KJ}$, and 
$\mathbb{I}_{ijkl} =\half (\delta_{ik}\delta_{jl} + \delta_{il}\delta_{jk})$.
Let $\alpha_I(\Bchi) = \alpha\Lambda_I(\Bchi)$, where $\Lambda_I = \sqrt{\sum_{i=1}^3\chi_{iI}^2}$ is the distortion of the crystal structure in the $\be_I$ direction due to the configurational change.
Also, let $\beta_{JK} = \beta$. Since $\partial \Lambda_I/\partial \Bchi = \Lambda_I^{-1}\Bchi\bM_I$, we have
\begin{align}
\frac{\partial \psi^\mathrm{S}}{\partial \Bchi}
 &=
\frac{1}{2} \sum_{I=1}^3 \frac{\alpha}{\Lambda_I}\Bchi\bM_I M_{II}^2
\end{align}

\subsubsection{Anisotropy induced by a configurational change in crystal structure}
\label{sec:anisotropy}

We consider changes in the material configuration that correspond to an evolution from a cubic crystal structure to three tetragonal crystal structures (in 3D), each oriented along one of the coordinate axes. The resulting anisotropy is reflected in the standard elastic deformation fields and the associated stresses. To demonstrate this effect, we consider two unit cubes, each initially with a cubic crystal structure. Through Dirichlet boundary conditions on the configurational domain, we force one cube into a tetragonal crystal structure oriented along $\be_1$ and the other cube into an $\be_2$-oriented tetragonal structure. Both cubes are also subjected to simple uniaxial tension along $\be_1$ through applied Dirichlet conditions on the standard displacement. The distinct stress plots of Figure~\ref{fig:anisotropy} show the resulting anisotropy. We used the three well free energy function for $\psi^\mathrm{M}$ and $\alpha_I(\Bchi) = \alpha(5\Lambda_I(\Bchi) - 4)$ to accentuate the anisotropy. The anisotropic tetragonal crystal structures in the two cases also produce distinct lateral deformation. Figure~\ref{fig:anisotropy2} compares the two computations, and the second case shows significantly less displacement in the $\be_2$ direction due to the $\be_2$-oriented tetragonal structure.

\begin{figure}[htb]
        \centering
\begin{minipage}[t]{0.35\textwidth}
        \centering
	\includegraphics[width=\textwidth]{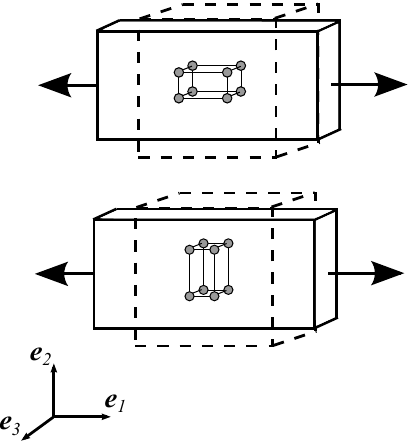}
\end{minipage}
\begin{minipage}[t]{0.6\textwidth}
        \centering
        \includegraphics[width=\textwidth]{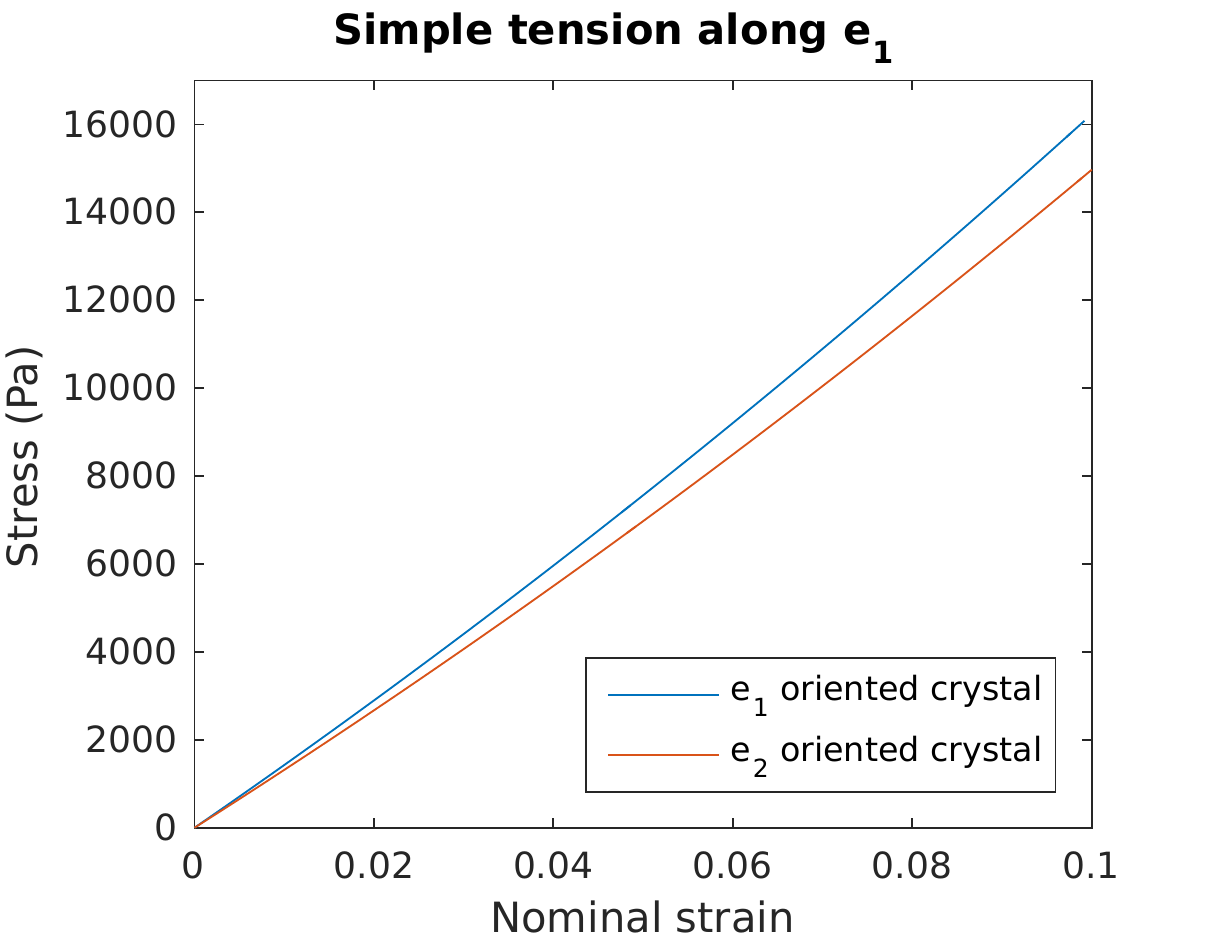}
\end{minipage}
        \caption{The $\be_1$-oriented tetragonal crystal structure leads to higher stresses than the $\be_2$-oriented tetragonal structure when subjected to simple uniaxial tension along $\be_1$. This demonstrates the differences in evolved anisotropy induced by the configurational changes in the two cases depicted on the left.}
	\label{fig:anisotropy}
\end{figure}

\begin{figure}[htb]
        \centering
        \includegraphics[width=\textwidth]{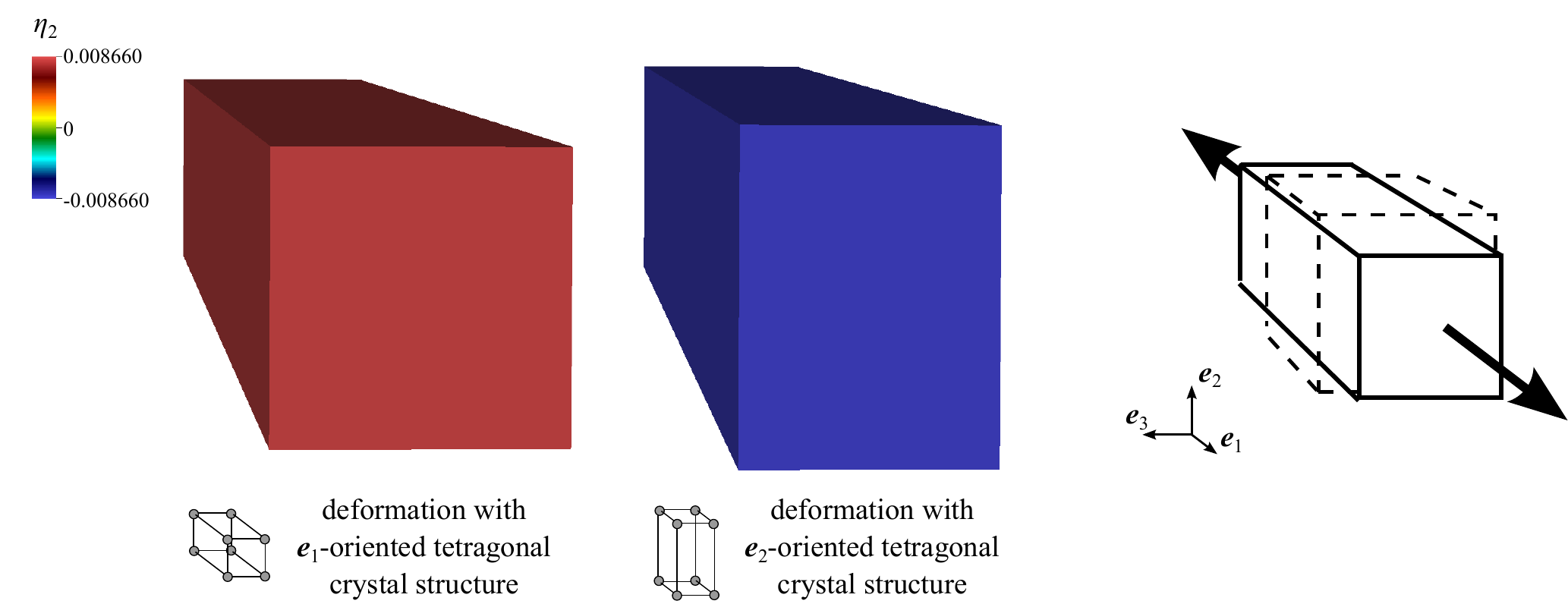}
        \caption{Computational results are compared for simple uniaxial tension along $\be_1$. The color contour plots of $\eta_2$ display the tetragonal variants, whose corresponding wells are located at $(\sqrt{3}/200,1/200)$, $(-\sqrt{3}/200,1/200)$, and $(0,-0.01)$ in $(\eta_2,\eta_3)$ space. The distortion has been scaled by $20\times$ the elastic deformation. The results on the right show less deformation in the $\be_2$ direction due to the anisotropy induced by the $\be_2$-oriented tetragonal crystal structure, compared to the case on the left.}
	\label{fig:anisotropy2}
\end{figure}

\subsection{Microstructure formation by evolution of the material configuration}

Figure \ref{fig:2Dbending} demonstrates a 2D plane strain problem wherein the material configuration evolves to distort the crystal structure from the the square to the rectangle. The beam is rapidly quenched from a high temperature causing the initially stable square structure to become unstable as the configuration-dependent component of the free energy function, $\psi^\mathrm{M}$, changes from convex to double-welled. The beam is then loaded in bending. The double-welled free energy renders the rectangular variants stable, and strain accommodation of the inhomogeneous configuration results in the microstructure shown. The parameter values used in equations \ref{eqn:psi_M} and \ref{eqn:psi_E} are $s = 0.1$, $d = 1$, $l = 0.1$ and $\mu = 1\times 10^{-1}$, $\beta = 1\times 10^{-1}$, $\alpha = 2\times 10^{-1}$, respectively. Contours of $\eta_2$ appear in the plots, where $\eta_2 = \pm 0.1$ locates the wells corresponding to the two rectangular variants, and $\eta_2 = 0$ is the square structure, which exists only in the interfaces between variants in this evolved material configuration. The fineness of the microstructure in the computations is determined by the gradient length scale parameter $l$. The configurational displacement $\bU$ at $x_1 = 10$ was specified as $0.5\bar{\bu}$. 

Figure \ref{fig:3Dbending} demonstrates the corresponding problem in 3D, where ``plane strain boundary conditions'' have been applied on the faces perpendicular to $\be_3$. The strain energy density function $\psi^\mathrm{M}$ allows for three tetragonal variants, but only two variants are seen due to the plane strain boundary conditions. The parameter values used in equations \ref{eqn:psi_M} and \ref{eqn:psi_E} are $s = 0.1$, $d = 1$, $l = 0.25$ and $\mu = 1\times 10^{-1}$, $\beta = 1\times 10^{-1}$, $\alpha = 2\times 10^{-1}$, respectively. Contours of $\eta_2$ appear in the plots, where the three tetragonal variants are located at $(\sqrt{3}/20,1/20)$, $(-\sqrt{3}/20,1/20)$, and $(0,-0.1)$ in $(\eta_2,\eta_3)$ space. Again, the configurational displacement $\bU$ at $x_1 = 10$ was specified as $0.5\bar{\bu}$. Note that a larger length scale parameter was used in the 3D problem, resulting in a coarser microstructure than the 2D bending problem.

      \begin{figure}[tb]
        \centering
        \includegraphics[width=\columnwidth]{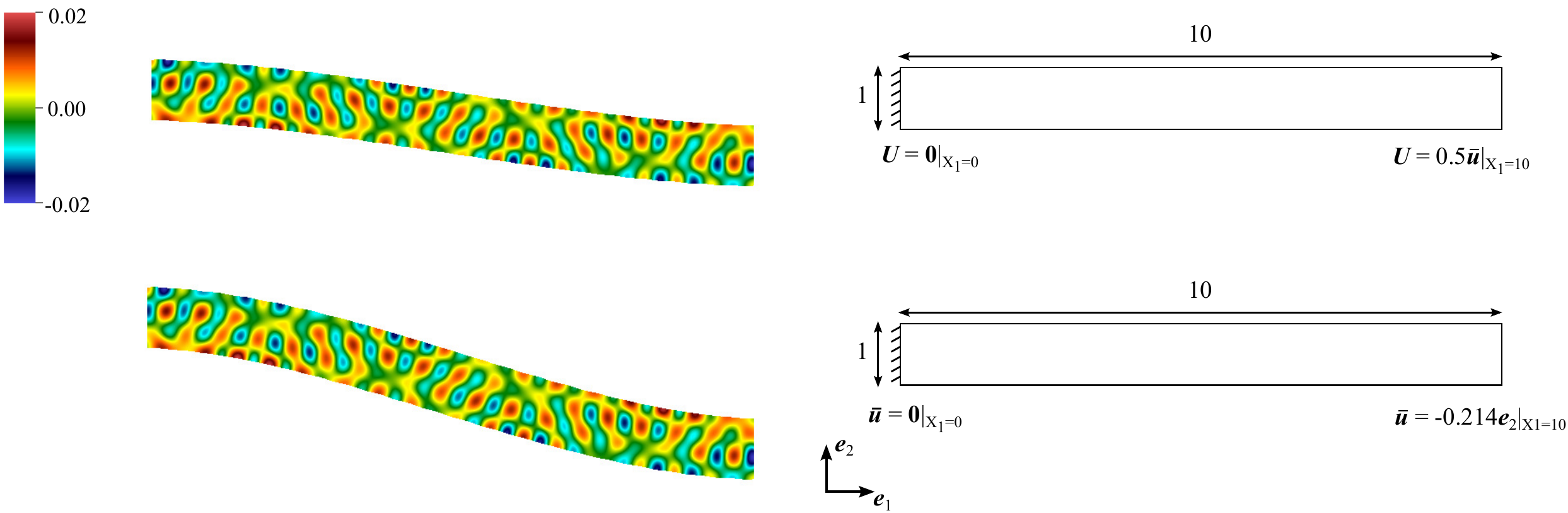}
	\caption{Simulation of 2D beam bending and the resulting material microstructure. The configurational displacement $\bU$ was specified to be $0.5\bar{\bu}$ at $X_2 = 10$. Contours of $\eta_2$ are plotted where the values $\pm 0.1$ locate the well corresponding to the two rectangular variants. The top plot is deformed by the configurational displacement and the bottom plot by the total displacement. The displacement for both plots is scaled by a factor of $10\times$.}
	\label{fig:2Dbending}
      \end{figure}

      \begin{figure}[tb]
        \centering
        \includegraphics[width=\columnwidth]{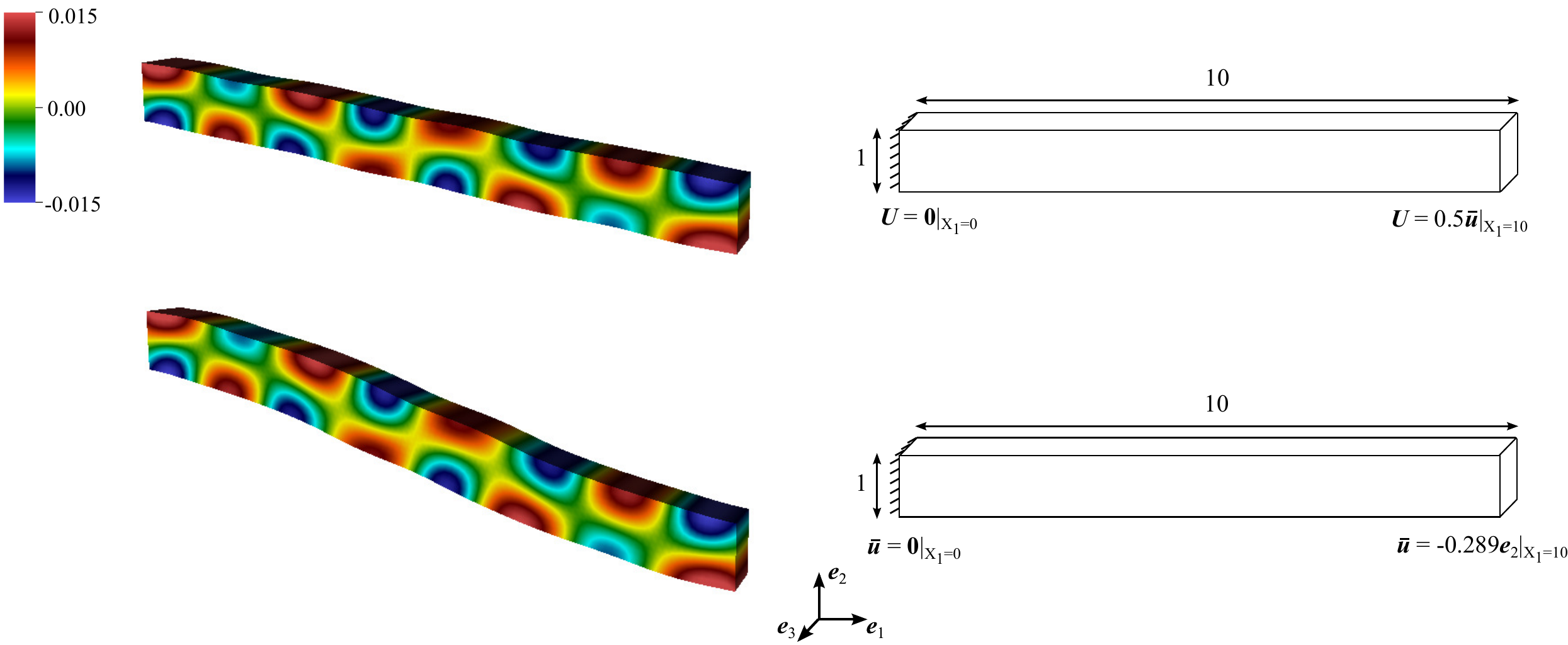}
	\caption{Simulation of 3D beam bending with plane strain boundary conditions and the resulting material microstructure. Contours of $\eta_2$ are plotted where the values $\pm 0.0866$ correspond to two of the three tetragonal variants. Only two variants are seen because of the plane strain boundary conditions. The top plot is deformed by the configurational displacement and the bottom plot by the total displacement. The displacement for both plots is scaled by a factor of $10\times$. A larger length scale parameter was used in the 3D problem than the 2D, resulting in a coarser microstructure.}
	\label{fig:3Dbending}
      \end{figure}

We note that the results of these computations compare well with those obtained by \citet{Rudrarajuetal2014}. However, in that work no configurational fields were identified. The entire problem was posed as a problem of elasticity relative to a high-symmetry (cubic or square) reference crystal. For a state where in the high-symmetry structure became unstable (by quenching, for instance) elastic deformation carried in the crystal structure into stable tetragonal states. The merits of the treatment presented here are that they allow us to separate out the configurational evolution as distinct from elastic deformation. This is particularly useful in describing anisotropy, as we have shown in Section \ref{sec:anisotropy}.

\pagebreak
\FloatBarrier

\section{Conclusion}
\label{sec:conclusion}
      \begin{figure}[htb]
        \centering
        \includegraphics[width=\textwidth]{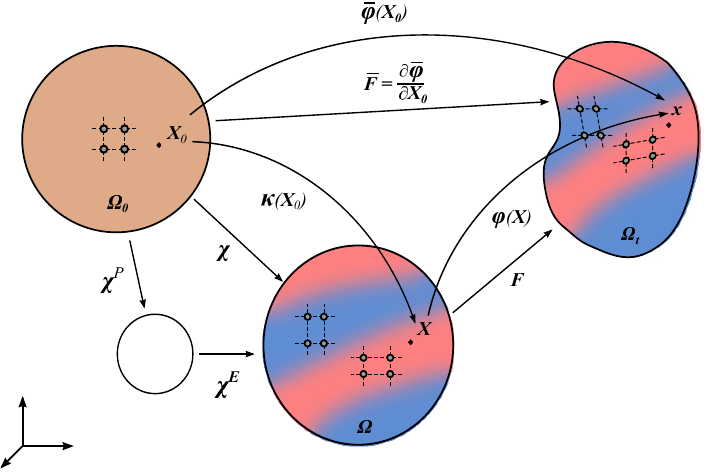}
	\label{fig:irreversibilities}
      \end{figure}

We have considered the modeling of materials with multiple solid phases within a continuum setting by separating the kinematics associated with the material evolution into a newly-identified configurational field, and the standard displacement field. This applies to interfacial phases changes maintaining a sharp interface and to volume phase changes resulting in multiple diffuse interfaces. By noting that the free energy density can be extended to depend on both these kinematic fields, we have obtained  equilibrium conditions associated with the configurational evolution, separately from those that hold for the standard displacement field. 

This separation of the deformation field into a configurational field in the material space and a standard spatial displacement field allows this framework to model a wide class of materials physics problems involving formation and movement of phase boundaries. In the context of crystalline materials, phase boundaries may occur due to nucleation and growth mechanisms, like those seen during precipitate evolution and formation of grain boundaries, or through phase transformations like martensitic transformations and twin-formation in HCP metals. All these phenomena involve sharp or diffused phase boundaries driven by interface kinetics or volumetric phase transformations. The framework presented here provides an overarching theoretical basis for representing the evolution of both sharp and diffuse phase boundaries. Notably, it also furnishes a variational basis for obtaining the governing partial differential equations.

In this first communication of these ideas, we have presented a preliminary exploration of phase transformations restricted to migrating sharp interfaces, such as arise at incoherent interphases, as well as phase transformations that occur throughout the volume of a material, resulting in diffuse interphase interfaces. We have shown that evolving elastic anisotropy due to the phase changes that distort the  crystal structure from a parent to a daughter phase can be captured through  a dependence of the free energy density function and of the conventional elastic moduli on the tangent map of the configurational field. Additionally, this formulation is able to reproduce the results previously obtained with a treatment of all deformation relative to a high symmetry reference crystal. 
 
Irreversibilities of crystallographic slip associated with the material evolution can be represented by imposing a further elasto-plastic decomposition on the tangent map of the configurational field $\Bchi = \Bchi^\mathrm{E}\Bchi^\mathrm{P}$, where $\Bchi^\mathrm{E}$ models the elastic distortion of the crystal structure and $\Bchi^\mathrm{P}$ models the crystallographic slip. Additionally, the common kinematic and variational underpinnings to the treatment of sharp and diffuse interfaces suggests the potential for modeling the evolution within a material from coherent (diffuse) to incoherent (sharp) interphase interfaces. The treatment introduced in this communication therefore has potential for modeling a wide array of phase transformations while clearly exposing the underlying configurational changes. That these configurational fields can be as diverse as that corresponding to interface motion in a non-crystalline material as well as crystal distortion is notable.

\section*{Acknowledgements}
The mathematical formulation for this work was carried out under an NSF DMREF grant: DMR1436154 ``DMREF: Integrated Computational Framework for Designing Dynamically Controlled Alloy-Oxide Heterostructures'', and an NSF CDI Type I grant: CHE1027729 ``Meta-Codes for Computational Kinetics''. GT was also partly supported by an NSF Graduate Research Fellowship under Grant No. DGE 1256260. The numerical formulation and computations have been carried out as part of research supported by the U.S. Department of Energy, Office of Basic Energy Sciences, Division of Materials Sciences and Engineering under Award \#DE-SC0008637 that funds the PRedictive Integrated Structural Materials Science (PRISMS) Center at University of Michigan. 

\pagebreak

\begin{appendix}

\section{Variational formulation for the sharp interface problem}
\label{sec:appa}
We consider variations on the configurational displacement, $\bU^\eps :=\bU + \eps \bW$, and on the Newtonian displacement, $\bu^\eps := \bu + \eps \bw$. We find the first variation using the functional defined over $\Omega_0$. At equilibrium, the first variation of the Gibbs free energy is zero.

\begin{align}
\frac{\mathrm{d}}{\mathrm{d}\eps} \Pi[\bu^\eps;\bU^\eps] \evat_{\eps=0}&
\begin{aligned}[t]
 =& 
\frac{\mathrm{d}}{\mathrm{d}\eps} \bigg\{ 
\int \limits_{{\Omega}_0}\psi(\bF^\eps,\Bkappa^\eps(\bX^0))\det\Bchi^\eps\,\mathrm{d}V_0\\
&
 - \int \limits_{{\Omega}_0} \bf(\Bkappa^\eps(\bX^0)) \cdot \bu^\eps\det\Bchi^\eps \, \mathrm{d}V_0\\
&
 -  \int \limits_{{\partial \Omega^\mathrm{S}_{T_0}}} \bT^0 \cdot \bu^\eps \, \mathrm{d}S_0 \bigg\}
\evat_{\eps=0}
\end{aligned} \nonumber \\
&
\begin{aligned}[t]
=& \bigg\{ 
\int \limits_{{\Omega}_0}\left[
\frac{\partial \psi}{\partial \bF}:\frac{d \bF^\eps}{d\eps}
+ \frac{\partial \psi}{\partial \Bkappa}\cdot\frac{d \Bkappa^\eps}{d\eps}
\right]\det\Bchi\,\mathrm{d}V_0\\
&+ \int \limits_{{\Omega}_0}\psi
\frac{\mathrm{d}}{\mathrm{d}\eps}\det\Bchi^\eps\,\mathrm{d}V_0
 - \int \limits_{{\Omega}_0} \bf\cdot\bu \frac{\mathrm{d}}{\mathrm{d}\eps}\det\Bchi^\eps\, \mathrm{d}V_0\\
&
 - \int \limits_{{\Omega}_0} \left( \bf\cdot \frac{d\bu^\eps}{d\eps}
+ \bu\cdot\frac{\partial \bf}{\partial \Bkappa} \cdot\frac{d\Bkappa^\eps}{d\eps}
\right) \det\Bchi\, \mathrm{d}V_0\\
&
 -  \int \limits_{{\partial \Omega^\mathrm{S}_{T_0}}} \bT^0 \cdot\frac{d\bu^\eps}{d\eps} \, \mathrm{d}S_0
\bigg\}\evat_{\eps=0}\\
\end{aligned} \nonumber \\
&=0
\label{eqn:PI_epsA}
\end{align}

Consider the first variation of $\bF$, recalling equation (\ref{eqn:F}):
\begin{align}
\frac{d \bF^\eps}{\partial \eps}
&\begin{aligned}[t]
=& \left(\frac{\partial \bw}{\partial \bX^0}
 +  \frac{\partial \bW}{\partial \bX^0}\right){\Bchi^\eps}^{-1}\\
& -
\left(\mathbbm{1} +  \frac{\partial\bu^\eps}{\partial \bX^0}
 +  \frac{\partial\bU^\eps}{\partial \bX^0}\right)
{\Bchi^\eps}^{-1} \frac{\partial \bW}{\partial \bX^0}{\Bchi^\eps}^{-1}
\end{aligned} \nonumber \\
&= \left(\frac{\partial \bw}{\partial \bX^0}
 +  \frac{\partial \bW}{\partial \bX^0}\right){\Bchi^\eps}^{-1} -
\bF^\eps \frac{\partial \bW}{\partial \bX^0}{\Bchi^\eps}^{-1} \nonumber \\
&= \left[ \frac{\partial \bw}{\partial \bX^0}
 + \left(\mathbbm{1} - \bF^\eps \right)
\frac{\partial \bW}{\partial \bX^0}
\right] {\Bchi^\eps}^{-1}
\label{eqn:F_epsA}
\end{align}

Now consider the first variation of $\det \Bchi$, recalling equation (\ref{eqn:chi}):
\begin{align}
\frac{d \det\Bchi^\eps}{d \eps} &=
\frac{\partial \det\Bchi^\eps}{\partial \Bchi^\eps}:\frac{d \Bchi^\eps}{d \eps} \nonumber \\
&= \det\Bchi^\eps {\Bchi^\eps}^{-T}:\frac{\partial \bW}{\partial \bX^0} \nonumber  \\
&=\iso:\left(\frac{\partial \bW}{\partial \bX^0} {\Bchi^\eps}^{-1}\right)\det\Bchi^\eps
\label{eqn:det_chi_epsA}
\end{align}

Substituting (\ref{eqn:F_epsA}) and (\ref{eqn:det_chi_epsA}) into (\ref{eqn:PI_epsA}) and using the relations $\partial \psi/\partial\bF = \bP$, $d\Bkappa^\eps/d\eps = \bW$, and $d\bu^\eps/d\eps = \bw$ gives

\begin{align}
\begin{aligned}[b]
0&=
\int \limits_{{\Omega}_0}\left[
\bP:
 \left[ \frac{\partial \bw}{\partial \bX^0} \Bchi^{-1}
 + \left(\mathbbm{1} - \bF \right)
\frac{\partial \bW}{\partial \bX^0} \Bchi^{-1}
\right]
+ \frac{\partial \psi}{\partial \Bkappa}\cdot\bW
\right]\det\Bchi\,\mathrm{d}V_0\\
&\phantom{=}
+ \int \limits_{{\Omega}_0} (\psi - \bf\cdot\bu)
\iso:\left(\frac{\partial \bW}{\partial \bX^0} \Bchi^{-1}\right)\det\Bchi
\,\mathrm{d}V_0\\
&\phantom{=}
 - \int \limits_{{\Omega}_0} \left( \bf\cdot \bw
+ \bu\cdot\frac{\partial \bf}{\partial \Bkappa} \cdot \bW
\right) \det\Bchi\, \mathrm{d}V_0
 -  \int \limits_{{\partial \Omega^\mathrm{S}_{T_0}}} \bT^0\cdot\bw\, \mathrm{d}S_0
\end{aligned}
\end{align}

We group terms according to $\bw$, $\bW$, and their gradients.

\begin{align}
\begin{aligned}[b]
0&= 
\int \limits_{{\Omega_0}}
\bP: \left(\frac{\partial \bw}{\partial \bX^0}\Bchi^{-1}\right)
\det\Bchi\,\mathrm{d}V_0\\
&\phantom{=}
 - \int \limits_{{\Omega_0}} (\bf\cdot\bw)\det\Bchi \, \mathrm{d}V_0
 -  \int \limits_{{\partial {\Omega}^\mathrm{S}_{T_0}}} \bT^0\cdot\bw \, \mathrm{d}S_0\\
&\phantom{=}
+ \int \limits_{{\Omega_0}}
\left(\bP - \left(\bf\cdot\bu\right)\iso + \Eshelby \right):
\left(\frac{\partial \bW}{\partial \bX^0}\Bchi^{-1}\right)
\det\Bchi\,\mathrm{d}V_0\\
&\phantom{=}
+ \int \limits_{{\Omega_0}}\left(
\frac{\partial\psi}{\partial \Bkappa}
-\left(\frac{\partial\bf}{\partial \Bkappa}\right)^T\bu
\right)\cdot\bW\det\Bchi\,\mathrm{d}V_0
\end{aligned}
\end{align}
Note that $\Eshelby:= \psi\iso - \bF^T\bP$ is the Eshelby stress tensor.
We now convert the integrals back to the $\Omega$ domain.
\begin{align}
\begin{aligned}[b]
0
&=
\int \limits_{{\Omega}}
\bP: \frac{\partial \bw}{\partial \bX}
\,\mathrm{d}V\\
&\phantom{=}
 - \int \limits_{{\Omega}} (\bf\cdot\bw) \, \mathrm{d}V
 -  \int \limits_{{ \partial {\Omega^\mathrm{S}_{T}}}} \bT\cdot\bw \, \mathrm{d}S\\
&\phantom{=}
+ \int \limits_{{\Omega}}
\left(\bP - \left(\bf\cdot\bu\right)\iso + \Eshelby \right):
\frac{\partial \bW}{\partial \bX}
\,\mathrm{d}V\\
&\phantom{=}
+ \int \limits_{{\Omega}}\left(
\frac{\partial\psi}{\partial \bX}
-\left(\frac{\partial \bf}{\partial \bX}\right)^T\bu
\right)\cdot\bW\,\mathrm{d}V
\end{aligned}
\end{align}
We now perform integration by parts, recognizing the potential jump terms at the interface of phases $\alpha$ and $\beta$.  We use the operator $\nabla\cdot$ here to refer to the divergence with respect to $\bX$. Note that $\bN$ is the unit normal to the boundary of the body, and $\bN^\Gamma$ is the normal to the interface.
\begin{align}
&\begin{aligned}[b]
0&= 
\int \limits_{{\Omega}}
[\nabla\cdot (\bw^T\bP) - \bw\cdot(\nabla\cdot \bP+\bf)]
\,\mathrm{d}V
-  \int \limits_{{ \partial {\Omega^\mathrm{S}_{T}}}} \bT\cdot\bw \, \mathrm{d}S\\
&\phantom{=}
+ \int \limits_{{\Omega}} \nabla\cdot\left[\bW^T
\left(\bP - \left(\bf\cdot\bu\right)\iso + \Eshelby \right)
\right]
\,\mathrm{d}V\\
&\phantom{=}
- \int \limits_{{\Omega}}
\bW\cdot\left[\nabla\cdot
\left(\bP - \left(\bf\cdot\bu\right)\iso + \Eshelby \right)
-\frac{\partial\psi}{\partial \bX}
+\left(\frac{\partial \bf}{\partial \bX}\right)^T\bu
\right]\,\mathrm{d}V
\end{aligned} \nonumber \\
&\phantom{0}\begin{aligned}[b]
&=
\int \limits_{{ \partial {\Omega^\mathrm{S}_{T}}}} \bw\cdot(\bP\bN-\bT) \, \mathrm{d}S
-\int \limits_{{\Omega}}
\bw\cdot(\nabla\cdot \bP+\bf)
\,\mathrm{d}V\\
&\phantom{=}
+\int \limits_{{\Gamma}}
\ljmp\bw\cdot\bP\bN^\Gamma
\rjmp\,\mathrm{d}S
+ \int \limits_{{\Gamma}} \ljmp\bW\cdot
\left(\bP - \left(\bf\cdot\bu\right)\iso + \Eshelby \right)
\bN^\Gamma\rjmp
\,\mathrm{d}S\\
&\phantom{=}
+ \int \limits_{ \partial {\Omega^\mathrm{M}_{T}}}\bW\cdot
\left[\bP - \left(\bf\cdot\bu\right)\iso + \Eshelby \right]
\bN
\,\mathrm{d}S  \\
&\phantom{=}
- \int \limits_{{\Omega}}
\bW\cdot\left[\nabla\cdot
\left(\bP - \left(\bf\cdot\bu\right)\iso + \Eshelby \right)
-\frac{\partial\psi}{\partial \bX}
+\left(\frac{\partial \bf}{\partial \bX}\right)^T\bu
\right]\,\mathrm{d}V
\end{aligned}
\end{align}
By allowing only continuous fields $\bW$ and $\bw$, we simplify further.
\begin{align}
\begin{aligned}[b]
0&=
\int \limits_{{ \partial {\Omega^\mathrm{S}_{T}}}} \bw\cdot(\bP\bN-\bT) \, \mathrm{d}S
-\int \limits_{{\Omega}}
\bw\cdot(\nabla\cdot \bP+\bf)
\,\mathrm{d}V\\
&\phantom{=}
+\int \limits_{{\Gamma}}
\bw\cdot\ljmp\bP\bN^\Gamma
\rjmp\,\mathrm{d}S
+ \int \limits_{{\Gamma}}\bW\cdot\ljmp
\left(\bP - \left(\bf\cdot\bu\right)\iso + \Eshelby \right)
\bN^\Gamma\rjmp
\,\mathrm{d}S\\
&\phantom{=}
+ \int \limits_{ \partial {\Omega^\mathrm{M}_{T}}}\bW\cdot
\left[\bP - \left(\bf\cdot\bu\right)\iso + \Eshelby \right]
\bN
\,\mathrm{d}S\\
&\phantom{=}
- \int \limits_{{\Omega}}
\bW\cdot\left[\nabla\cdot
\left(\bP - \left(\bf\cdot\bu\right)\iso + \Eshelby \right)
-\frac{\partial\psi}{\partial \bX}
+\left(\frac{\partial \bf}{\partial \bX}\right)^T\bu
\right]\,\mathrm{d}V
\end{aligned}
\end{align}
The corresponding strong form for the sharp interface problem consists of the two following sets of equations. The second set of equations (\ref{eqn:confBCA} - \ref{eqn:confPDEA}) has been simplified under the assumption that the first set of equations (\ref{eqn:standBCA} - \ref{eqn:standPDEA}) is satisfied.
\begin{subequations}
\begin{align}
\bP\bN - \bT &=0 \text{ on } \partial\Omega^\mathrm{S}_{T}
\label{eqn:standBCA}\\
\ljmp \bP \bN^\Gamma \rjmp &= 0 \text{ on } \Gamma\\
\nabla\cdot \bP + \bf &= 0 \text{ in } \Omega
\label{eqn:standPDEA}\\
\left(\Eshelby + \bP - (\bf\cdot\bu)\iso\right)\bN &=0 \text{ on } \partial\Omega^\mathrm{M}_{T}
\label{eqn:confBCA}\\
\ljmp \left(\Eshelby - (\bf\cdot\bu)\iso\right)\bN^\Gamma \rjmp &= 0 \text{ on } \Gamma\\
\nabla\cdot \Eshelby 
-\frac{\partial\psi}{\partial \bX} 
- \bF^T\bf &= 0 \text{ in } \Omega
\label{eqn:confPDEA}
\end{align}
\end{subequations}
Consider the following simplification of equation (\ref{eqn:confPDEA}):

\begin{align}
0 &=
\nabla\cdot \left(\Eshelby - (\bf\cdot\bu)\iso\right) - \bf
-\frac{\partial\psi}{\partial \bX} 
+\left(\frac{\partial \bf}{\partial \bX}\right)^T\bu \nonumber \\
&=
 \nabla\cdot \Eshelby - \left(\frac{\partial \bf}{\partial \bX}\right)^T\bu - \left(\frac{\partial \bu}{\partial \bX} +\iso\right)^T\bf
-\frac{\partial\psi}{\partial \bX} 
+\left(\frac{\partial \bf}{\partial \bX}\right)^T\bu \nonumber \\
&= \nabla\cdot \Eshelby 
-\frac{\partial\psi}{\partial \bX}  - \bF^T\bf
\end{align}

\section{First variation of constant interfacial energy}
\label{sec:appc}
The mean curvature-driven term in Equation (\ref{eqn:elastpluscurvdom}) and the additional boundary condition (\ref{eqn:elastpluscurvboun}) can be obtained by considering pure curvature-driven motion. We seek to minimize the interface energy
\begin{align}
\Pi^\Gamma &= \int_\Gamma \psi^\Gamma\,\mathrm{d}S
\end{align}
with respect to the interface $\Gamma$, where $\psi^\Gamma$ is a constant. To do so, we find $\Gamma$ such that the first variation of $\Pi$ is zero. We define the interface $\Gamma$ by the parameterization $\br(u,v)$, where $u$ and $v$ are defined over the domain $T$.  We vary $\br(u,v)$ by $\eps \bW$ to allow variations of the interface location. To avoid integration over a varying surface, we perform a change of variables. We let $\Gamma^0$ be a surface defined by the parameterization $\br^0(u,v)$ where $\br^0$, $\br$, and $\br^\eps$ are related as follows:
\begin{align}
\br(u,v) &= \br^0(u,v) + \bU(\br^0(u,v))\\
\br^\eps(u,v) &= \br^0(u,v) + \bU(\br^0(u,v)) + \eps \bW(\br^0(u,v))\label{eqn:r_eps}
\end{align}
Then we can write
\begin{align}
 \int_\Gamma \psi^\Gamma\,\mathrm{d}S &=
  \int_T \psi^\Gamma\left|\br_{,u}\times\br_{,v}\right|\,\mathrm{d}u\mathrm{d}v
\end{align}
The first variation is
\begin{align}
\frac{\mathrm{d}}{\mathrm{d}\eps}\Pi^{\Gamma_\eps}\big|_{\eps=0} &=  
\frac{\mathrm{d}}{\mathrm{d}\eps}
\left\{
\int_{\Gamma^\eps} \psi^\Gamma\,\mathrm{d}S^\eps
\right\} \Big|_{\eps=0}\nonumber\\
&=
\frac{\mathrm{d}}{\mathrm{d}\eps}
\left\{
\int_T \psi^\Gamma\left|\br^\eps_{,u}\times\br^\eps_{,v}\right|\,\mathrm{d}u\mathrm{d}v
\right\} \Big|_{\eps=0}\nonumber\\
&=
\int_{T} \psi^\Gamma
\frac{\tfrac{\mathrm{d}}{\mathrm{d}\eps}
\left(\br^\eps_{,u}\times\br^\eps_{,v}\right)\big|_{\eps=0}
\cdot\left(\br_{,u}\times\br_{,v}\right)}
{\left|\br_{,u}\times\br_{,v}\right|}\,\mathrm{d}u\mathrm{d}v\nonumber\\
&=
\int_{T} \psi^\Gamma
{\tfrac{\mathrm{d}}{\mathrm{d}\eps}
\left(\br^\eps_{,u}\times\br^\eps_{,v}\right)\big|_{\eps=0}
\cdot\bN^\Gamma}
\,\mathrm{d}u\mathrm{d}v \label{eqn:genIntegral}
\end{align}
From equation (\ref{eqn:r_eps}), we have
\begin{align}
\tfrac{\mathrm{d}}{\mathrm{d}\eps} \br^\eps \big|_{\eps=0}
&= \bW 
\end{align}
Substituting this result gives
\begin{align}
\frac{\mathrm{d}}{\mathrm{d}\eps}\Pi^{\Gamma_\eps}\big|_{\eps=0}
 &=  
\int_{T} \psi^\Gamma
{\left(\bW_{,u}\times\br_{,v} + \br_{,u}\times\bW_{,v}\right)
\cdot\bN^\Gamma} \,\mathrm{d}u\mathrm{d}v\nonumber\\
 &=  
 \int_{T} \psi^\Gamma
{\left[((\nabla\bW)\br_{,u})\times\br_{,v} - ((\nabla\bW)\br_{,v})\times\br_{,u}\right]
\cdot\bN^\Gamma} \,\mathrm{d}u\mathrm{d}v
\end{align}
Continuing in coordinate notation for clarity, this becomes
\begin{align}
\frac{\mathrm{d}}{\mathrm{d}\eps}\Pi^{\Gamma_\eps}\big|_{\eps=0}
 &=  
  \int_{T} \psi^\Gamma \partial_\ell(W_j)
\left(r_{\ell,u}r_{k,v} - r_{\ell,v}r_{k,u}\right)
\eps_{ijk}N_i^\Gamma \,\mathrm{d}u\mathrm{d}v\nonumber\\
 &=  
  \int_{T} \psi^\Gamma \partial_\ell(W_j)
\left(\delta_{\ell m}\delta_{kn} - \delta_{\ell n}\delta_{km}\right)
r_{m,u}r_{n,v}
\eps_{ijk}N_i^\Gamma \,\mathrm{d}u\mathrm{d}v\nonumber\\
 &=  
  \int_{T} \psi^\Gamma \partial_\ell(W_j)
\eps_{p\ell k}\eps_{pmn}
r_{m,u}r_{n,v}
\eps_{ijk}N_i^\Gamma \,\mathrm{d}u\mathrm{d}v
\end{align}
Note that 
\begin{align}
\eps_{pmn}
r_{m,u}r_{n,v} &= (\br_{,u}\times\br_{,v})_p\nonumber\\
&= N^\Gamma_p|\br_u\times\br_v|
\end{align}
Using this result lets us write the integral over $\Gamma$.
\begin{align}
\frac{\mathrm{d}}{\mathrm{d}\eps}\Pi^{\Gamma_\eps}\big|_{\eps=0}
 &=  
   \int_{T} \psi^\Gamma \partial_\ell(W_j)
\eps_{p\ell k}
\eps_{ijk}N_i^\Gamma N^\Gamma_p|\br_u\times\br_v|
 \,\mathrm{d}u\mathrm{d}v\nonumber\\
  &=  
   \int_{\Gamma} \psi^\Gamma \partial_\ell(W_j)
\eps_{p\ell k}
\eps_{ijk}N_i^\Gamma N^\Gamma_p
 \,\mathrm{d}S\nonumber\\
  &=  
   \int_{\Gamma} \psi^\Gamma 
\left[
\partial_\ell(
N_i^\Gamma W_j
\eps_{ijk})
-
\partial_\ell(N_i^\Gamma )
W_j
\eps_{ijk}
\right]
\eps_{p\ell k}
N^\Gamma_p
 \,\mathrm{d}S\nonumber\\
   &=  
   \int_{\Gamma} \psi^\Gamma 
[
\partial_\ell(
N_i^\Gamma W_j
\eps_{ijk})
\eps_{p\ell k}
N^\Gamma_p
-
\partial_\ell(N_i^\Gamma )
W_j
(\delta_{ip}\delta_{j\ell} -
\delta_{i\ell}\delta_{jp})
N^\Gamma_p]
 \,\mathrm{d}S\nonumber\\
   &=  
   \int_{\Gamma} \psi^\Gamma 
[
\partial_\ell(
N_i^\Gamma W_j
\eps_{ijk})
\eps_{p\ell k}
N^\Gamma_p
-
\partial_j(N_i^\Gamma )
N^\Gamma_i
W_j
+
\partial_i(N_i^\Gamma )
W_j
N^\Gamma_j]
 \,\mathrm{d}S
\end{align}
The term $\partial_j(N_i^\Gamma )N^\Gamma_i$ reduces to zero since $N_i^\Gamma N_i^\Gamma = 1$. The result can be expressed in direct notation.

\begin{align}
\frac{\mathrm{d}}{\mathrm{d}\eps}\Pi^{\Gamma_\eps}\big|_{\eps=0}
 &=  
 \int_{\Gamma} \psi^\Gamma
 \left[
 \nabla\times\left(
 \bN^\Gamma\times\bW\right)
 +
\left(\nabla\cdot
 \bN^\Gamma\right)
 \bW
 \right]\cdot\bN^\Gamma
 \,\mathrm{d}S
 \end{align}
 We apply Stoke's theorem to the first term and use $\nabla\cdot \bN^\Gamma = -2H$ , where $H$ is the mean curvature.
 \begin{align}
\frac{\mathrm{d}}{\mathrm{d}\eps}\Pi^{\Gamma_\eps}\big|_{\eps=0}
 &=  
 \int_{\Gamma} -2\psi^\Gamma H
\left(
 \bW
 \cdot\bN^\Gamma
\right) \,\mathrm{d}S
 +
\oint_{\partial \Gamma}
\psi^\Gamma(\bN^\Gamma\times\bW)
\cdot\mathrm{d}\br
 \end{align}

At equilibrium, the first variation of the total energy is zero, giving the following result.

 \begin{align}
0
 &=  
 \int_{\Gamma} -2\psi^\Gamma H
\left(
 \bW
 \cdot\bN^\Gamma
\right) \,\mathrm{d}S
 +
\oint_{\partial \Gamma}
\psi^\Gamma(\bN^\Gamma\times\bW)
\cdot\mathrm{d}\br \label{eqn:gamma_zero}
 \end{align}

If we have a full Dirichlet condition on the boundary of $\Gamma$ then $\bW = 0$ on $\partial\Gamma$ and the line integral is equal to zero. This gives the result
\begin{align}
0 &=  
 \int_{\Gamma} -2\psi^\Gamma H
 \left( \bW
 \cdot\bN^\Gamma
\right)
\,\mathrm{d}S
\end{align}
The other possible condition is to allow the boundary of the interface $\Gamma$ to move within the boundary $\partial \Omega$. This corresponds to $\bU\cdot\bN = \bW\cdot\bN = 0$, where $\bN$ is the outward unit normal to $\partial\Omega$. We will designate a boundary with this condition as $\partial\Gamma_T$. On such boundaries, we can express $\bW$ using the orthonormal basis $\{\bN,\bT^\Gamma,\bN\times\bT^\Gamma\}$, where $\bT^\Gamma$ is the unit tangent vector to $\partial\Gamma$. Then we have
\begin{align}
\bW &= W^T\bT^\Gamma + W^{N\times T}\left(\bN\times\bT^\Gamma\right)
\end{align}
which satisfies the condition $\bW\cdot\bN = 0$. We substitute this expression into the line integral:
\begin{align}
\oint_{\partial \Gamma}
\psi^\Gamma(\bN^\Gamma\times\bW)
\cdot\mathrm{d}\br
&=
\oint_{\partial \Gamma}
\psi^\Gamma\left[
\bN^\Gamma\times
\left[
W^T\bT^\Gamma + W^{N\times T}\left(\bN\times\bT^\Gamma\right)
\right]
\right]
\cdot\mathrm{d}\br
 \end{align}
 Since $\bT^\Gamma$ and $\mathrm{d}\br$ have the same orientation, $(\bN^\Gamma\times\bT^\Gamma)\cdot\mathrm{d}\br = 0$. We also use the identity $\ba\times(\bb\times\bc) = \bb(\ba\cdot\bc) - \bc(\ba\cdot\bb)$.
 \begin{align}
\oint_{\partial \Gamma}
\psi^\Gamma(\bN^\Gamma\times\bW)
\cdot\mathrm{d}\br
&=
\oint_{\partial \Gamma}
\psi^\Gamma W^{N\times T}\left[
\bN(\bN^\Gamma\cdot\bT^\Gamma) - \bT^\Gamma(\bN^\Gamma\cdot\bN)
\right]
\cdot\mathrm{d}\br\nonumber\\
&=
-\oint_{\partial \Gamma}
\psi^\Gamma W^{N\times T} (\bN^\Gamma\cdot\bN)
\bT^\Gamma\cdot\mathrm{d}\br
 \end{align}
 since $\bN^\Gamma$ and $\bT^\Gamma$ are orthogonal. Finally, we use $W^{N\times T} = \bW\cdot\left(\bN\times\bT^\Gamma\right)$ and substitute into equation (\ref{eqn:gamma_zero}) to get the complete condition for equilibrium with respect to interfacial energy.
 \begin{align}
0
 &=  
 \int_{\Gamma} -2\psi^\Gamma H
\left(
 \bW
 \cdot\bN^\Gamma
\right) \,\mathrm{d}S
-\oint_{\partial \Gamma}
\psi^\Gamma \bW\cdot\left(\bN\times\bT^\Gamma\right) (\bN^\Gamma\cdot\bN)
\bT^\Gamma\cdot\mathrm{d}\br
 \end{align}
 Note that $\bT^\Gamma\cdot\mathrm{d}\br$ and $\bN\times\bT^\Gamma$ are always nonzero, so we get the following strong form.
\begin{subequations}
 \begin{align}
 -2\psi^\Gamma H &= 0  \text{ in } \Gamma  \label{eqn:interfacedom}\\
 \psi^\Gamma\bN^\Gamma\cdot\bN &= 0 \text{ on } \partial\Gamma_T\label{eqn:interface}
 \end{align} 
\end{subequations}
If the interfacial energy is nonzero, this requires zero mean curvature within the interface and that the interface is perpendicular to the boundary of the body where they meet. If the interfacial energy $\psi^\Gamma$ provides only one contribution to the driving force on $\Gamma$, as in Section \ref{sec:interfaceEnergy}, then the left hand-side of Equation (\ref{eqn:interfacedom}) is the corresponding contribution to Equation (\ref{eqn:elastpluscurvdom}), while (\ref{eqn:interface}) is the additional boundary condition (\ref{eqn:elastpluscurvboun}).\\
 
\pagebreak

\section{Variational formulation for the diffuse interface problem}
\label{sec:appb}

Let $\bU^\eps := \bU + \eps \bW$ and $\bar{\bu}^\eps := \bar{\bu} + \eps \bar{\bw}$. Then, recalling equation \ref{eqn:diffuseGibbs}, equilibrium requires the following:
\begin{align}
\begin{aligned}[b]
\frac{\mathrm{d}}{\mathrm{d}\eps} \Pi[\bar{\bu}^\eps;\bU^\eps] \evat_{\eps=0}
&= 
\frac{\mathrm{d}}{\mathrm{d}\eps} \bigg\{ 
\int \limits_{{\Omega}_0} \psi^\mathrm{M}(\bX^0,\Bchi^\eps,\nabla^0\Bchi^\eps) \,\mathrm{d}V_0\\
&\phantom{=}
+ \int \limits_{{\Omega}_0}
\psi^\mathrm{S}(\bX^\eps,\bF^\eps,\Bchi^\eps)\det\Bchi^\eps \,\mathrm{d}V_0\\
&\phantom{=}
  - \int \limits_{{\Omega}_0} \bf^0\cdot \bar{\bu}^\eps \, \mathrm{d}V_0 -  
\int \limits_{ \partial {\Omega^\mathrm{S}_{T_0}}} \bT^0 \cdot \bar{\bu}^\eps \, \mathrm{d}S_0 \bigg\}\evat_{\eps=0}\\
&= 0
\end{aligned}
\end{align}
Then
\begin{align}
\begin{aligned}[b]
0
&= 
\int \limits_{{\Omega}_0} \left(
 \frac{\partial \psi^\mathrm{M}}{\partial \Bchi}:
\frac{\mathrm{d} \Bchi^\eps}{\mathrm{d} \eps} \evat_{\eps=0}
+ \frac{\partial \psi^\mathrm{M}}{\partial \nabla^0\Bchi}\vdots
\frac{\mathrm{d} \nabla^0\Bchi^\eps}{\mathrm{d} \eps} \evat_{\eps=0}
\right) \, \mathrm{d}V_0\\ 
&\phantom{=}
+ \int \limits_{{\Omega}_0}
\left( \frac{\partial \psi^\mathrm{S}}{\partial \bX}\cdot\frac{d\bX^\eps}{d \eps} \evat_{\eps=0}
+ \frac{\partial \psi^\mathrm{S}}{\partial \bF}:\frac{d\bF^\eps}{d \eps}\evat_{\eps=0}
 \right) \det\Bchi \, \mathrm{d}V_0\\
&\phantom{=}
+ \int \limits_{{\Omega}_0}
\left( 
\frac{\partial \psi^\mathrm{S}}{\partial \Bchi}:\frac{d \Bchi^\eps}{d \eps} \evat_{\eps=0}
\det\Bchi
+
\psi^\mathrm{S} \frac{d \det\Bchi^\eps}{d \eps}\evat_{\eps=0}
\right) \, \mathrm{d}V_0\\
&\phantom{=}
- \int \limits_{{\Omega}_0}
\bf^0\cdot  \bar{\bw} \, \mathrm{d}V_0
 -  \int \limits_{ \partial {\Omega^\mathrm{S}_{T_0}}} \bT^0\cdot\bar{w}  \, \mathrm{d}S_0
\end{aligned}
\end{align}

We apply the earlier results concerning the first variations of $\bF$ and $\det\Bchi$. We also define
$\bB :=  \partial \psi^\mathrm{M}/\partial \nabla^0\Bchi$
 and $J_\chi := \det\Bchi$. 
Then we have
\begin{align}
\begin{aligned}[b]
0 &=
\int \limits_{{\Omega}_0} \left(
\frac{\partial \psi^\mathrm{M}}{\partial \Bchi}:
\nabla^0 \bW 
+ \bB\vdots\nabla^0\nabla^0\bW
\right) \, \mathrm{d}V_0\\
& \phantom{=}
+
\int \limits_{{\Omega}_0}
 \bigg(
\frac{\partial \psi^\mathrm{S}}{\partial \bX}\cdot \bW
+ \bP: \left[\left(  \nabla^0\bar{\bw} - 
\bF \nabla^0\bW
\right)\Bchi^{-1}
\right]
  \bigg) J_\chi \, \mathrm{d}V_0\\
&\phantom{=}
+ \int \limits_{{\Omega}_0} 
 \left[ \frac{\partial \psi^\mathrm{S}}{\partial \Bchi}:
 \nabla^0\bW 
+\psi^\mathrm{S}\iso:
 \left(\nabla^0\bW \Bchi^{-1}\right)\right]
 J_\chi\, \mathrm{d}V_0\\
&\phantom{=}
 - \int \limits_{{\Omega}_0}
 \bf^0\cdot \bar{\bw} \, \mathrm{d}V_0
 -  \int \limits_{ \partial {\Omega^\mathrm{S}_{T_0}}} \bT^0\cdot \bar{\bw} \, \mathrm{d}S_0
\end{aligned}
\end{align}

We group terms by $\bar{\bw}$, $\bW$, and their gradients and use the Eshelby stress tensor, $\mathcal{E} := \psi^\mathrm{S}\mathbbm{1} - \bF^T\bP$. The resulting weak form is as follows:
\begin{align}
\begin{aligned}[b]
0 &=
\int \limits_{{\Omega}_0}
\bB\vdots\nabla^0\nabla^0\bW \, \mathrm{d}V_0
+
\int \limits_{{\Omega}_0}
J_\chi\frac{\partial \psi^\mathrm{S}}{\partial\bX}\cdot \bW \, \mathrm{d}V_0\\
&\phantom{=}
+\int \limits_{{\Omega}_0}
 \left[\frac{\partial \psi^\mathrm{M}}{\partial \Bchi} + 
J_\chi\left(\Eshelby \Bchi^{-T}
 + \frac{\partial \psi^\mathrm{S}}{\partial \Bchi} \right) \right]:
 \nabla^0\bW \, \mathrm{d}V_0\\
&\phantom{=}
+
\int \limits_{{\Omega}_0}
J_\chi \left( \bP \Bchi^{-T} \right):
 \nabla^0\bar{\bw} \, \mathrm{d}V_0
- 
\int \limits_{{\Omega}_0}
 \bf^0\cdot  \bar{\bw} \, \mathrm{d}V_0
 -  
\int \limits_{\partial \Omega^\mathrm{S}_{T_0}} \bT^0\cdot \bar{\bw} \, \mathrm{d}S_0
\end{aligned}
\end{align}

Applying integration by parts gives the following result.
\begin{align}
\begin{aligned}[b]
0 &=
\int \limits_{{\Omega}_0}
\bB\vdots \nabla^0\nabla^0\bW \, \mathrm{d}V_0\\
&\phantom{=}
+\int \limits_{ \partial {\Omega^\mathrm{M}_{T_0}}}
 \bW\cdot\left[\frac{\partial \psi^\mathrm{M}}{\partial \Bchi} + 
J_{\chi}\left(\Eshelby \Bchi^{-T}
 + \frac{\partial \psi^\mathrm{S}}{\partial \Bchi} \right)
\right] \bN^0 \, \mathrm{d}S_0\\
&\phantom{=}
- \int \limits_{{\Omega}_0}  \bW\cdot \left(\nabla^0\cdot
\left[\frac{\partial \psi^\mathrm{M}}{\partial \Bchi} + 
J_{\chi}\left( \Eshelby  \Bchi^{-T}
 + \frac{\partial \psi^\mathrm{S}}{\partial \Bchi} \right) 
\right]
- J_{\chi}\frac{\partial \psi^\mathrm{S}}{\partial \bX}
\right) \, \mathrm{d}V_0\\
& \phantom{=}
+
\int \limits_{ \partial {\Omega^\mathrm{S}_{T_0}}}   \bar{\bw}\cdot
\left( J_{\chi} \bP \Bchi^{-T} \bN^0
- \bT^0 \right)
 \, \mathrm{d}S_0\\
&\phantom{=}
-
\int \limits_{{\Omega}_0} \bar{\bw}\cdot\left[\nabla^0\cdot
\left(J_{\chi} \bP \Bchi^{-T} \right)
+  \bf^0
 \right] \, \mathrm{d}V_0
\end{aligned}
\label{eqn:intbyparts}
\end{align}

Deriving the strong form from this weak form involves several additional terms due to the dependence on $\nabla^0 \Bchi$, as described in \citet{Rudrarajuetal2014}.
We use the normal and surface gradient operators,  $\nabla^n$ and $\nabla^s$, where
\begin{align}
\nabla^n \psi &= \nabla^0\psi\cdot\bN^0\\
\nabla^s \psi &= \nabla^0 \psi - \left(\nabla^n\psi\right)\bN^0
\end{align}
Also, $\bb = -\nabla^s \bN^0 = \bb^T$
is the second fundamental form of the smooth parts of the boundary, $\partial\Omega_0$ and $\bN^E = \BXi\times\bN^0$, where $\BXi$ is the unit tangent to the smooth curve $\mathcal{C}_0$ that forms an edge between subsets $\partial \Omega^+_0$ and $\partial \Omega^-_0$of the smooth boundary surfaces $\partial \Omega_0$. If $\bN^{\mathcal{C}^+}$ is the outward unit normal to $\mathcal{C}_0$ from $\partial \Omega^+_0$ and $\bN^{\mathcal{C}^-}$ is the outward unit normal to $\mathcal{C}_0$ from $\partial \Omega^-_0$, then we define $\ljmp \bB:\left( \bN^\mathcal{C} \otimes \bN^0\right)  \rjmp^\mathcal{C} := \bB:\left( \bN^{\mathcal{C}^+} \otimes \bN^0\right) + \bB:\left( \bN^{\mathcal{C}^-} \otimes \bN^0\right)$.
We can then write
\begin{align}
\begin{aligned}[b]
\int \limits_{{\Omega}_0}
\bB\vdots \nabla^0\nabla^0\bW \, \mathrm{d}V_0
&=
\int \limits_{{\Omega}_0}  \bW\cdot\nabla^0\nabla^0\bB \, \mathrm{d}V_0
- \int \limits_{\partial \Omega_{0}}  \bW\cdot \bC \, \mathrm{d}S_0\\
&\phantom{=}
+ \int \limits_{\mathcal{C}_0} \bW\cdot
\ljmp \bB:(\bN^\mathcal{C}\otimes \bN^0 \rjmp^\mathcal{C}
\, \mathrm{d}L_0\\
&\phantom{=}
+ \int \limits_{\partial \Omega_{0}}
\nabla^n \bW\cdot \bB:\left(\bN^0\otimes \bN^0\right) \, \mathrm{d}S_0
\end{aligned}
\end{align}

where, using coordinate notation for clarity,
\begin{align}
\begin{aligned}[b]
C_I &= \nabla^nB_{I\gamma\zeta}N^0_\zeta N^0_\gamma
+2\nabla^s_\gamma B_{I\gamma\zeta} N^0_\zeta\\
&\phantom{=}
+B_{I\gamma\zeta}\nabla^s_\gamma N^0_\zeta
- (b_{\xi\xi} N^0_\gamma N^0_\zeta - b_{\gamma\zeta})B_{I\gamma\zeta}
\end{aligned}
\end{align}

Applying this result to equation \ref{eqn:intbyparts} gives the following:

\begin{align}
\begin{aligned}[b]
0 &=
\int \limits_{{\Omega}_0}  \bW\cdot\nabla^0\nabla^0\bB \, \mathrm{d}V_0
+ \int \limits_{\mathcal{C}_0} \bW\cdot
\ljmp \bB:\left( \bN^\mathcal{C} \otimes \bN^0\right)  \rjmp^\mathcal{C}
\, dL_0\\
&\phantom{=}
- \int \limits_{\partial \Omega^\mathrm{M}_{T_0}}  \bW\cdot\bC \, \mathrm{d}S_0
+ \int \limits_{\partial \Omega^\mathrm{M}_{T_0}}
\nabla^n \bW\cdot \bB:\left(\bN^0\otimes \bN^0\right) \, \mathrm{d}S_0
\\
&\phantom{=}
+
\int \limits_{\partial \Omega^\mathrm{M}_{T_0}}
 \bW\cdot\left[\frac{\partial \psi^\mathrm{M}}{\partial \Bchi} + 
J_{\chi}\left(\Eshelby\Bchi^{-T}
 + \frac{\partial \psi^\mathrm{S}}{\partial \Bchi} \right)
\right] \bN^0 \, \mathrm{d}S_0\\
&\phantom{=}
- \int \limits_{{\Omega}_0}  \bW\cdot \left(
\nabla^0\cdot\left[\frac{\partial \psi^\mathrm{M}}{\partial \Bchi} + 
J_{\chi}\left(\Eshelby \Bchi^{-T}
 + \frac{\partial \psi^\mathrm{S}}{\partial \Bchi} \right) 
\right]
- J_{\chi}\frac{\partial \psi^\mathrm{S}}{\partial \bX}
\right) \, \mathrm{d}V_0\\
&\phantom{=}
 +
\int \limits_{ \partial {\Omega^\mathrm{S}_{T_0}}}   \bar{\bw}\cdot\left(
J_{\chi} \bP \Bchi^{-T} \bN^0
- \bT^0 \right)
 \, \mathrm{d}S_0\\
&\phantom{=}
-
\int \limits_{{\Omega}_0} \bar{\bw}\cdot\left[
\nabla^0\cdot \left(J_{\chi} \bP \Bchi^{-T} \right)
+  \bf^0
 \right] \, \mathrm{d}V_0
\end{aligned}
\end{align}

Applying the appropriate integration by parts and standard variational arguments leads to the following strong form.

\begin{subequations}
\begin{align}
J_{\chi} \bP \Bchi^{-T} \bN^0
- \bT^0  &= 0 \text{ on } \partial {\Omega^\mathrm{M}_{T_0}}\\
\nabla^0\cdot \left(J_{\chi} \bP \Bchi^{-T} \right)
+  \bf^0 &= 0 \text{ in } \Omega_0
\label{eqn:Bstdpde}\\
\ljmp \bB:\left( \bN^\mathcal{C} \otimes \bN^0\right)  \rjmp^\mathcal{C}
&= 0  \text{ on }  \mathcal{C}^\mathrm{M}_{T_0}\\
\bB:\left( \bN^0 \otimes \bN^0\right)
&= 0  \text{ on } \partial {\Omega^\mathrm{S}_{T_0}}\\
\frac{\partial \psi^\mathrm{M}}{\partial \Bchi}\bN^0 + 
J_{\chi}\left(\Eshelby\Bchi^{-T}
 + \frac{\partial \psi^\mathrm{S}}{\partial \Bchi} \right) \bN^0
-\bC &= 0  \text{ on } \partial {\Omega^\mathrm{S}_{T_0}}\label{eqn:Bconfigtracbc}\\
\nabla^0\cdot \left[\frac{\partial \psi^\mathrm{M}}{\partial \Bchi} + 
J_{\chi}\left(\Eshelby \Bchi^{-T}
 + \frac{\partial \psi^\mathrm{S}}{\partial \Bchi} \right) 
\right]
- J_{\chi}\frac{\partial \psi^\mathrm{S}}{\partial \bX}
 -  \nabla^0\nabla^0\bB &= 0 \text{ in } \Omega_0\label{eqn:Bconfigpde}
\end{align}
\end{subequations}

Consider the simplification of equation \ref{eqn:Bconfigpde}, using coordinate notation for clarity:
\begin{align}
0 &= \left[\frac{\partial \psi^\mathrm{M}}{\partial \chi_{I\alpha}} + 
J_{\chi}\left(\mathcal{E}_{IJ} \chi^{-1}_{\alpha J}
 + \frac{\partial \psi^\mathrm{S}}{\partial \chi_{I\alpha}} \right) 
\right]_{,\alpha}
- J_{\chi}\frac{\partial \psi^\mathrm{S}}{\partial X_I}
 -  B_{I\alpha\beta,\alpha\beta} \nonumber \\
\phantom{0}&
\begin{aligned}[b]
 &= \left(\frac{\partial \psi^\mathrm{M}}{\partial \chi_{I\alpha}}\right)_{,\alpha}
 + 
\left(J_{\chi}\psi^\mathrm{S} \chi^{-1}_{\alpha I}\right)_{,\alpha}
 -
\left(J_{\chi} F_{iI}P_{iJ}\chi^{-1}_{\alpha J}\right)_{,\alpha}
 + 
\left(J_{\chi}\frac{\partial \psi^\mathrm{S}}{\partial \chi_{I\alpha}} \right)_{,\alpha}\\
&\phantom{=}
- J_{\chi}\frac{\partial \psi^\mathrm{S}}{\partial X_I}
 -  B_{I\alpha\beta,\alpha\beta}
\end{aligned} \nonumber \\
\phantom{0}&
\begin{aligned}[b]
 &= \left(\frac{\partial \psi^\mathrm{M}}{\partial \chi_{I\alpha}}\right)_{,\alpha}
 + 
J_{\chi}\left(\frac{\partial \psi^\mathrm{S}}{\partial X^0_\alpha}+
\frac{\partial \psi^\mathrm{S}}{\partial F_{iJ}}F_{iJ,\alpha}\right) \chi^{-1}_{\alpha I}
 + 
\psi^\mathrm{S}\left(J_{\chi} \chi^{-1}_{\alpha I}\right)_{,\alpha}\\
&\phantom{=}
 -
F_{iI,\alpha}\left(J_{\chi} P_{iJ}\chi^{-1}_{\alpha J}\right)
-
F_{iI}\left(J_{\chi} P_{iJ}\chi^{-1}_{\alpha J}\right)_{,\alpha}
 + 
\left(J_{\chi}\frac{\partial \psi^\mathrm{S}}{\partial \chi_{I\alpha}} \right)_{,\alpha}
- J_{\chi}\frac{\partial \psi^\mathrm{S}}{\partial X_I}
 -  B_{I\alpha\beta,\alpha\beta}
\end{aligned} \nonumber \\
\phantom{0}&
\begin{aligned}[b]
 &= \left(\frac{\partial \psi^\mathrm{M}}{\partial \chi_{I\alpha}}\right)_{,\alpha}
 + 
J_{\chi}P_{iJ}F_{iJ,I}
 + 
\psi^\mathrm{S}\left(J_{\chi} \chi^{-1}_{\alpha I}\right)_{,\alpha}\\
&\phantom{=}
 -
J_{\chi}F_{iI,J}P_{iJ}
-
F_{iI}\left(J_{\chi} P_{iJ}\chi^{-1}_{I\alpha}\right)_{,\alpha}
 + 
\left(J_{\chi}\frac{\partial \psi^\mathrm{S}}{\partial \chi_{I\alpha}} \right)_{,\alpha}
 -  
B_{I\alpha\beta,\alpha\beta} \label{eqn:Bconfigpde_sim}
\end{aligned}
\end{align}
Consider the term $\left(J_{\chi} \chi^{-1}_{\alpha I}\right)_{,\alpha}$ and use the relation $\chi_{J\beta,\alpha} = \chi_{J\alpha,\beta}$
\begin{align}
\left(J_{\chi} \chi^{-1}_{\alpha I}\right)_{,\alpha} &=
\frac{\partial \left(J_{\chi} \chi^{-1}_{\alpha I}\right)}{\partial \chi_{J\beta}}\chi_{J\beta,\alpha} \nonumber \\
&=
\left(\frac{\partial J_{\chi} }{\partial \chi_{J\beta}}\chi^{-1}_{\alpha I}
+ J_{\chi}\frac{\partial  \chi^{-1}_{\alpha I}}{\partial \chi_{J\beta}}\right)\chi_{J\beta,\alpha} \nonumber \\
&=
\left(J_\chi \chi^{-1}_{\beta J}\chi^{-1}_{\alpha I}
- J_{\chi}\chi^{-1}_{\alpha J}\chi^{-1}_{\beta I}\right)\chi_{J\beta,\alpha} \nonumber\\
&= 0
\end{align}
Substitute this result, the relation $F_{iI,J} = F_{iJ,I}$, and equation \ref{eqn:Bstdpde} into equation \ref{eqn:Bconfigpde_sim}. This gives
\begin{align}
 \left(\frac{\partial \psi^\mathrm{M}}{\partial \chi_{I\alpha}}
 + 
J_{\chi}\frac{\partial \psi^\mathrm{S}}{\partial \chi_{I\alpha}} \right)_{,\alpha}
 +
F_{iI}f^0_i -  B_{I\alpha\beta,\alpha\beta}
&= 0
\end{align}
or, in direct notation,
\begin{align}
\nabla^0\cdot
 \left(\frac{\partial \psi^\mathrm{M}}{\partial \Bchi}
 + 
J_{\chi}\frac{\partial \psi^\mathrm{S}}{\partial \Bchi} \right)
 +
\bF^T\bf^0 - \nabla^0\nabla^0\bB
&= 0
\end{align}

\end{appendix}

%
%
\section*{References}
\bibliographystyle{abbrvnat}
\bibliography{references}

\begin{thebibliography}{32}
\providecommand{\natexlab}[1]{#1}
\providecommand{\url}[1]{\texttt{#1}}
\expandafter\ifx\csname urlstyle\endcsname\relax
  \providecommand{\doi}[1]{doi: #1}\else
  \providecommand{\doi}{doi: \begingroup \urlstyle{rm}\Url}\fi

\bibitem[Acharya and Fressengeas(2012)]{AcharyaFressengeas2012}
A.~Acharya and C.~Fressengeas.
\newblock Coupled phase transformations and plasticity as a field theory of
  deformation incompatibility.
\newblock \emph{International Journal of Fracture}, 74:\penalty0 87--94, 2012.
\newblock \doi{10.1007/s10704-011-9656-0}.

\bibitem[Ball and Crooks(2011)]{BallCrooks2011}
J.~M. Ball and E.~C.~M. Crooks.
\newblock Local minimizers and planar interfaces in a phase-transition model
  with interfacial energy.
\newblock \emph{Calculus of Variations}, 40:\penalty0 501--538, 2011.

\bibitem[Barth and Sethian(1998)]{BarthSethian1998}
T.~J. Barth and J.~A. Sethian.
\newblock Numerical schemes for the {Hamilton-Jacobi} and level set equations
  on triangulated domains.
\newblock \emph{J. Comput. Phys.}, 145:\penalty0 1--40, 1998.

\bibitem[Bhattacharya and Kohn(1997)]{BhattacharyaKohn1997}
K.~Bhattacharya and R.~V. Kohn.
\newblock Elastic energy minimization and the recoverable strains of
  polycrystalline shape-memory materials.
\newblock \emph{Archive for Rational Mechanics and Analysis}, 139:\penalty0
  99--180, 1997.

\bibitem[Bhattacharya et~al.(2004)Bhattacharya, Conti, Zanzotto, and
  Zimmer]{Bhattacharyaetal2004}
K.~Bhattacharya, S.~Conti, G.~Zanzotto, and J.~Zimmer.
\newblock Crystal symmetry and the reversibility of martensitic
  transformations.
\newblock \emph{Nature}, 428:\penalty0 55--59, 2004.

\bibitem[Brooks and Hughes(1982)]{BrooksHughes1982}
A.~N. Brooks and T.~J.~R. Hughes.
\newblock Streamline upwind/{Petrov-Galerkin} formulations for convection
  dominated flows with particular emphasis on the incompressible
  {Navier-Stokes} equations.
\newblock \emph{Comput. Methods Appl. Mech. Eng.}, 32:\penalty0 199--259, 1982.

\bibitem[Burton(1892)]{Burton1892}
C.~V. Burton.
\newblock A theory concerning the constitution of matter.
\newblock \emph{Philos. Mag.}, 33:\penalty0 191--204, 1892.

\bibitem[Denzer and Menzel(2014)]{DenzerMenzel2014}
R.~Denzer and A.~Menzel.
\newblock Configurational forces for quasi-incompressible large strain
  electro-viscoelasticity e application to fracture mechanics.
\newblock \emph{European Journal of Mechanics A/Solids}, 48:\penalty0 3--15,
  2014.

\bibitem[Eshelby(1951)]{Eshelby1951}
J.~D. Eshelby.
\newblock The force on an elastic singularity.
\newblock \emph{Philos. Trans. R. Soc. London, Ser. A}, 244\penalty0
  (877):\penalty0 87--112, Nov 1951.
\newblock URL \url{http://rsta.royalsocietypublishing.org/content/244/877/87}.

\bibitem[Garikipati and Rao(2001)]{GarikipatiRao2001}
K.~Garikipati and V.~S. Rao.
\newblock Recent advances in models for thermal oxidation of silicon.
\newblock \emph{J. Comput. Phys.}, 174\penalty0 (1):\penalty0 138--170,
  November 2001.

\bibitem[Garikipati et~al.(2006)Garikipati, Olberding, Narayanan, Arruda,
  Grosh, and Calve]{Garikipatietal2006}
K.~Garikipati, J.~E. Olberding, H.~Narayanan, E.~M. Arruda, K.~Grosh, and
  S.~Calve.
\newblock Biological remodelling: Stationary energy, configurational change,
  internal variables and dissipation.
\newblock \emph{J. Mech. Phys. Solids}, 54:\penalty0 1493--1515, 2006.

\bibitem[Gurtin(2000)]{Gurtin2000}
M.~E. Gurtin.
\newblock \emph{Configurational forces as basic concepts of continuum physics}.
\newblock New York: Springer, 2000.

\bibitem[Kalpakides and Arvanitakis(2009)]{KalpakidesArvanitakis2009}
V.~K. Kalpakides and A.~I. Arvanitakis.
\newblock Configurational forces in continuous theories of elastic
  ferroelectrics.
\newblock In P.~Steinmann, editor, \emph{IUTAM Symposium on Progress in the
  Theory and Numerics of Configurational Mechanics}, volume~17 of \emph{IUTAM
  Bookseries}, pages 229--238. Springer Netherlands, 2009.
\newblock URL \url{http://dx.doi.org/10.1007/978-90-481-3447-2_21}.

\bibitem[Kienzler and Herrmann(1997)]{KienzlerHerrmann1997}
R.~Kienzler and G.~Herrmann.
\newblock On the properties of the {Eshelby} tensor.
\newblock \emph{Acta Mech.}, 125\penalty0 (1-4):\penalty0 73--91, March 1997.
\newblock URL \url{http://link.springer.com/article/10.1007%2FBF01177300}.

\bibitem[Larmor(1897)]{Larmor1897}
J.~Larmor.
\newblock A dynamical theory of the electric and luminiferous medium - iii.
  relations with material media.
\newblock \emph{Philos. Trans. R. Soc. London, Ser. A}, 190:\penalty0 205--300,
  1897.

\bibitem[Macklin and Lowengrub(2006)]{MacklinLowengrub2006}
P.~Macklin and J.~Lowengrub.
\newblock An improved geometry-aware curvature discretization for level set
  methods: Application to tumor growth.
\newblock \emph{J. Comput. Phys.}, 215:\penalty0 392--401, 2006.

\bibitem[Maugin(1995)]{Maugin1995}
G.~A. Maugin.
\newblock Material forces: Concepts and applications.
\newblock \emph{Applied Mechanics Reviews}, 48\penalty0 (5):\penalty0 213--245,
  1995.
\newblock URL \url{http://dx.doi.org/10.1115/1.300510}.

\bibitem[Maugin(2011)]{Maugin2011}
G.~A. Maugin.
\newblock \emph{Configurational forces : thermomechanics, physics, mathematics,
  and numerics}.
\newblock Boca Raton, FL: Chapman \& Hall/CRC, 2011.

\bibitem[Mueller and Maugin(2002)]{MullerMaugin2002}
R.~Mueller and G.~A. Maugin.
\newblock On material forces and finite element discretizations.
\newblock \emph{Computational Mechanics}, 29\penalty0 (1):\penalty0 52--60,
  July 2002.
\newblock URL \url{http://link.springer.com/article/10.1007/s00466-002-0322-2}.

\bibitem[M{\"u}ller(1999)]{Muller1999}
S.~M{\"u}ller.
\newblock \emph{Calculus of Variations and Geometric Evolution Problems:
  Lectures given at the 2nd Session of the Centro Internazionale Matematico
  Estivo (C.I.M.E.) held in Cetraro, Italy, June 15--22, 1996}, chapter
  Variational models for microstructure and phase transitions, pages 85--210.
\newblock Springer Berlin Heidelberg, Berlin, Heidelberg, 1999.

\bibitem[Osher and Sethian(1988)]{OsherSethian1988}
S.~Osher and J.~A. Sethian.
\newblock Fronts propagating with curvature dependent speed: algorithms based
  on {Hamilton-Jacobi} formulation.
\newblock \emph{J. Comput. Phys.}, 79\penalty0 (12):\penalty0 12--49, 1988.

\bibitem[Podio-Guidugli(2002)]{PodioGuidugli2002}
P.~Podio-Guidugli.
\newblock Configurational forces: are they needed?
\newblock \emph{Mech. Res. Commun.}, 29\penalty0 (6):\penalty0 513--519,
  November-December 2002.

\bibitem[Rao and Hughes(2000)]{RaoHughes2000}
V.~S. Rao and T.~J.~R. Hughes.
\newblock On modelling thermal oxidation of silicon i: theory.
\newblock \emph{Int. J. Numer. Methods Eng.}, 47\penalty0 (1-3):\penalty0
  341--358, 2000.
\newblock ISSN 1097-0207.
\newblock URL
  \url{http://dx.doi.org/10.1002/(SICI)1097-0207(20000110/30)47:1/3<341::AID-NME774>3.0.CO;2-Z}.

\bibitem[Rao et~al.(2000)Rao, Hughes, and Garikipati]{Raoetal2000}
V.~S. Rao, T.~J.~R. Hughes, and K.~Garikipati.
\newblock On modelling thermal oxidation of silicon ii: numerical aspects.
\newblock \emph{Int. J. Numer. Methods Eng.}, 47\penalty0 (1-3):\penalty0
  359--377, 2000.
\newblock ISSN 1097-0207.
\newblock URL
  \url{http://dx.doi.org/10.1002/(SICI)1097-0207(20000110/30)47:1/3<359::AID-NME775>3.0.CO;2-7}.

\bibitem[Rudraraju et~al.(2014)Rudraraju, der Ven, and
  Garikipati]{Rudrarajuetal2014}
S.~Rudraraju, A.~V. der Ven, and K.~Garikipati.
\newblock Three-dimensional isogeometric solutions to general boundary value
  problems of {Toupin's} gradient elasticity theory at finite strains.
\newblock \emph{Comput. Methods Appl. Mech. Eng.}, 278:\penalty0 705--728,
  2014.

\bibitem[Rudraraju et~al.(2016)Rudraraju, der Ven, and
  Garikipati]{Rudrarajuetal2015}
S.~Rudraraju, A.~V. der Ven, and K.~Garikipati.
\newblock Mechano-chemical spinodal decomposition: {A} phenomenological theory
  of phase transformations in multi-component crystalline solids.
\newblock \emph{Nature npj Computational Materials}, 2, 2016.
\newblock URL \url{http://dx.doi.org/10.1038/npjcompumats.2016.12}.

\bibitem[Russo and Smereka(2000)]{RussoSmereka2000}
G.~Russo and P.~Smereka.
\newblock A remark on computing distance functions.
\newblock \emph{J. Comput. Phys.}, 163:\penalty0 51--67, 2000.

\bibitem[Steinmann(2002)]{Steinmann2002}
P.~Steinmann.
\newblock On spatial and material settings of hyperelastostatic crystal
  defects.
\newblock \emph{J. Mech. Phys. Solids}, 50:\penalty0 1743--1766, 2002.

\bibitem[Toupin(1962)]{Toupin1962}
R.~Toupin.
\newblock Elastic materials with couple-stresses.
\newblock \emph{Arch. Ration. Mech. Anal.}, 11:\penalty0 385--414, 1962.

\bibitem[Vu and Steinmann(2012)]{VuSteinmann2012}
D.~Vu and P.~Steinmann.
\newblock On the spatial and material motion problems in nonlinear
  electro-elastostatics with consideration of free space.
\newblock \emph{Mathematics and Mechanics of Solids}, 17\penalty0 (8):\penalty0
  803--823, 2012.

\bibitem[Yavari and Goriely(2013)]{YavariGoriely2013}
A.~Yavari and A.~Goriely.
\newblock Nonlinear elastic inclusions in isotropic solids.
\newblock \emph{Proceedings of the Royal Society of London A: Mathematical,
  Physical and Engineering Sciences}, 469\penalty0 (2160), 2013.
\newblock ISSN 1364-5021.
\newblock \doi{10.1098/rspa.2013.0415}.
\newblock URL
  \url{http://rspa.royalsocietypublishing.org/content/469/2160/20130415}.

\bibitem[Yavari et~al.(2006)Yavari, Marsden, and Ortiz]{Yavarietal2006}
A.~Yavari, J.~E. Marsden, and M.~Ortiz.
\newblock On spatial and material covariant balance laws in elasticity.
\newblock \emph{Journal of Mathematical Physics}, 47:\penalty0 042903, 2006.

\end{thebibliography}

\end{document}